\documentclass[trackchanges,twocolumn]{aastex701}
\usepackage{bm}
\usepackage{xspace}
\usepackage{threeparttable}
\usepackage{amsmath}
\usepackage{gensymb}
\usepackage{textcomp}
\DeclareUnicodeCharacter{2212}{\textminus}

\newcommand{\CST}{\texttt{CST}\xspace}
\newcommand{\STU}{\texttt{ST-U}\xspace}
\newcommand{\STUEL}{\texttt{ST-U+EL}\xspace}
\newcommand{\STUPL}{\texttt{ST-U+PL}\xspace}
\newcommand{\STCST}{\texttt{ST+CST}\xspace}
\newcommand{\STCDT}{\texttt{ST+CDT}\xspace}
\newcommand{\STPST}{\texttt{ST+PST}\xspace}
\newcommand{\STPDT}{\texttt{ST+PDT}\xspace}

\newcommand{\Msun}{$M_{\odot}$\xspace}
\newcommand{\Req}{$R_{\rm eq}$\xspace}
\newcommand{\jsf}{J1614\xspace}
\newcommand{\jsflong}{PSR~J1614$-$2230\xspace}
\newcommand{\josft}{PSR~J0614$-$3329\xspace}
\newcommand{\josfy}{PSR~J0740$+$6620\xspace}
\newcommand{\jofts}{PSR~J0437$-$4715\xspace}
\newcommand{\jdoty}{PSR~J0030$+$0451\xspace}
\newcommand{\nh}{$N_{\rm H}$\xspace}

\graphicspath{{./}{figures/}}

\begin{document}

\title{A NICER view of \jsflong: a massive and compact millisecond pulsar}

\author[0000-0002-3408-2759]{Lucien~Mauviard}
\affil{Universit\'{e} de Toulouse, CNES, CNRS, IRAP, Toulouse, France}
\email[show]{lucien.mauviard@utoulouse.fr}

\author[0000-0002-6449-106X]{Sebastien~Guillot}
\affil{Universit\'{e} de Toulouse, CNES, CNRS, IRAP, Toulouse, France}
\email{sebastien.guillot@utoulouse.fr}

\author[0000-0003-3783-7448]{Lami Suleiman}
\affil{Deutsches Elektronen-Synchrotron DESY, Platanenallee 6, 15738 Zeuthen, Germany}
\affil{Deutsches Zentrum für Astrophysik (DZA), Postplatz 1, 02826 Görlitz, Germany}
\email{lami.suleiman@desy.de}

\author[0000-0002-0428-8430]{Yves~Kini}
\affil{Gravitation and Astroparticle Physics Amsterdam (GRAPPA), University of Amsterdam, 1098XH Amsterdam, The Netherlands}
\email{y.kini@uva.nl}

\author[0000-0001-5848-0180]{Denis~Gonz\'alez-Caniulef}
\affil{Universit\'{e} de Toulouse, CNES, CNRS, IRAP, Toulouse, France}
\email{}

\author[0009-0008-3894-4783]{Christine~Kazantsev}
\affil{Universit\'{e} de Toulouse, CNES, CNRS, IRAP, Toulouse, France}
\email{}

\author[0009-0005-7766-5638]{Pierre~Stammler}
\affil{Universit\'{e} de Toulouse, CNES, CNRS, IRAP, Toulouse, France}
\email{}

\author[0000-0002-2651-5286]{Devarshi~Choudhury}
\affil{Anton Pannekoek Institute for Astronomy, University of Amsterdam, Science Park 904, 1098XH Amsterdam, the Netherlands}
\email{}

\author[0000-0002-9407-0733]{Bas~Dorsman}
\affil{Anton Pannekoek Institute for Astronomy, University of Amsterdam, Science Park 904, 1098XH Amsterdam, the Netherlands}
\email{}

\author[0009-0005-8019-0426]{Mariska~Hoogkamer}
\affiliation{Anton Pannekoek Institute for Astronomy, University of Amsterdam, Science Park 904, 1098XH Amsterdam, the Netherlands}
\email{}

\author[0000-0002-1169-7486]{Daniela~Huppenkothen}
\affiliation{Anton Pannekoek Institute for Astronomy, University of Amsterdam, Science Park 904, 1098XH Amsterdam, the Netherlands}
\email{}

\author[0000-0001-6356-125X ]{Tuomo~Salmi}
\affil{Department of Physics, University of Helsinki, P.O. Box 64, FI-00014 University of Helsinki, Finland}
\email{}

\author[0000-0002-1009-2354]{Anna~L.~Watts}
\affil{Gravitation and Astroparticle Physics Amsterdam (GRAPPA), University of Amsterdam, 1098XH Amsterdam, The Netherlands}
\affil{Anton Pannekoek Institute for Astronomy, University of Amsterdam, Science Park 904, 1098XH Amsterdam, the Netherlands}
\email{}

\author[0000-0002-6029-4712]{J\'{e}r\^{o}me~Novak}
\affil{Observatoire astronomique de Strasbourg, CNRS, Universit\'{e} de Strasbourg, 11 rue de l’Universit\'{e}, 67000 Strasbourg, France}
\affil{LUX, CNRS UMR 8262, Observatoire de Paris–PSL, Sorbonne Universit\'{e} Paris, 5 place Jules Janssen, 92190 Meudon, France}
\email{}

\author[0000-0002-2956-6062]{Natalie~A.~Webb}
\affil{Universit\'{e} de Toulouse, CNES, CNRS, IRAP, Toulouse, France}
\email{}

\author{Jean-Francois~Olive}
\affil{Universit\'{e} de Toulouse, CNES, CNRS, IRAP, Toulouse, France}
\email{}

\author[0000-0002-0893-4073]{Matthew~Kerr}
\affil{Space Science Division, U.S. Naval Research Laboratory, Washington, DC 20375, USA}
\email{}

\author[0000-0002-9049-8716]{Lucas~Guillemot}
\affil{LPC2E, OSUC, Universit\'{e} d’Orl\'{e}ans, CNRS, CNES, Observatoire de Paris, F-45071 Orl\'{e}ans, France}
\affil{Observatoire Radioastronomique de Nançay, Observatoire de Paris, Universit\'{e} PSL, Universit\'{e} d’Orl\'{e}ans, CNRS, 18330 Nançay, France}
\email{}

\author[0000-0002-1775-9692]{Isma\"{e}l~Cognard}
\affil{LPC2E, OSUC, Universit\'{e} d’Orl\'{e}ans, CNRS, CNES, Observatoire de Paris, F-45071 Orl\'{e}ans, France}
\affil{Observatoire Radioastronomique de Nançay, Observatoire de Paris, Universit\'{e} PSL, Universit\'{e} d’Orl\'{e}ans, CNRS, 18330 Nançay, France}
\email{}

\author[0000-0002-3649-276X]{Gilles~Theureau}
\affil{LPC2E, OSUC, Universit\'{e} d’Orl\'{e}ans, CNRS, CNES, Observatoire de Paris, F-45071 Orl\'{e}ans, France}
\affil{Observatoire Radioastronomique de Nançay, Observatoire de Paris, Universit\'{e} PSL, Universit\'{e} d’Orl\'{e}ans, CNRS, 18330 Nançay, France}
\affil{Laboratoire Univers et Th\'{e}ories LUTh, Observatoire de Paris, Universit\'{e} PSL, CNRS, Universit\'{e} de Paris, 92190 Meudon, France}
\email{}

\begin{abstract}
Using pulse profile modeling, we obtain the mass-radius measurement of a millisecond pulsar (MSP) with data from the Neutron Star Interior Composition ExploreR, XMM-Newton and the Chandra X-ray Observatory. We report here the radius of \jsflong, the second most massive MSP confirmed by radio timing. All of the data sets are well described by a simple model composed of two circular hot spots. The final result yields an equatorial radius of $R_{\rm eq}=10.06 ^{+1.25}_{-0.87}$~km, and a gravitational mass of $M=1.937^{+0.012}_{-0.013}\,$\Msun (equally tailed 68\% credible intervals). Although a non-thermal component was previously reported at higher energies, we find no sign of it in either our phase-averaged or phase-resolved spectral analyses. Using new relations linking the compactness to oblateness or surface gravity, and tailored to the spin frequency of \jsflong, we infer a configuration with one hot spot near the pole, and another near the equator. The tight mass posterior is essentially informed by radio timing, while the radius constraint is not as tight due to the low source signal (8.5$\sigma$ X-ray pulse significance). However, over all geometries and atmosphere models tested, the radius posterior tends toward low values ($\lesssim12.08$\,km, 90th percentile in all cases).
\end{abstract}


\section{Introduction}
\label{sec:introduction}

Neutron stars (NSs) are ultra dense objects (up to $\rho \sim 10^{15}\,\rm{g/cm}^3$ in their center) that prove to be unique laboratories to study the physics of dense matter. The matter in their core reaches densities of several times the nuclear saturation density. The properties of such matter, which has yet to be constrained, is encoded in macroscopic observables of NS, such as mass, radius, tidal deformability or moment of inertia. The measurement of these quantities is crucial to constrain dense matter.

Millisecond pulsars (MSPs) are NSs that have been spun up to rotation periods of a few milliseconds by the continued accretion of a companion star \citep{Alpar+82,Radhakrishnan+82}. Due to the combination of their fast spin and substantial magnetic field ($\sim 10^8$\,G), electron/positron pairs are created in their magnetosphere, accelerated and channeled onto their magnetic poles \citep{Ruderman+75}. These electron/positron showers heat the surface of the MSP, resulting in hot spots with temperatures of $\sim 10^6$\,K, emitting electromagnetic radiation in the soft X-rays \citep{Aarons+81,Harding+01,Harding+02}.

Modeling the energy-resolved surface emission of these hot spots as the MSP rotate, a method coined pulse profile modeling (PPM), proves to be a useful tool to inform on their radius. This is partially due to special and general relativistic effects, both imprinting their signature on the electromagnetic emission. Observing and quantifying these effects allow to probe their rotation speed and compactness, which in turn help to constrain the radius of MSPs \citep{Pavlov+97,Zavlin+97,Zavlin+02,Poutanen+03,Poutanen08,Bogdanov+19_II,Bogdanov+21_III}. For spin frequencies above 300~Hz, such modeling needs to take into account the oblate shape of the star to be sufficiently accurate \citep{Cadeau+07,Morsink+07}. The oblate shape of the star depends on its spin frequency, compactness, and only weakly on the nature of the dense matter inside it \citep{AGM14}.

The Neutron Star Interior Composition ExploreR (NICER, \citealt{NICER}) is an X-ray instrument built to measure these effects in order to quantify masses and radii of MSPs with PPM. For this purpose, it features a timing resolution sufficient to observe millisecond period pulses. Six MSPs have had their mass and radius measured with NICER data \citep{Riley+19, Miller+19, Riley+21, Miller+21, Salmi+22, Salmi+24_J0740, Dittmann+24, Choudhury+24_J0437, Vinciguerra+24_J0030, Salmi+24_J1231, Hoogkamer+25, Mauviard+25, Miller+26, Kini+26, Gonzalez-Caniulef+26} and we present here the application of the same method in order to yield a mass-radius measurement for a new MSP.

PPM radius measurements provide valuable input on the equation of state (EoS) of cold dense matter in the core of NSs. Alongside mass measurements with radio observations and gravitational waves from double NS mergers, they narrow down the space of possible EoS (see, e.g., \citealt{Lattimer12,Baym+18,Raaijmakers+21,Pang+21,Chatziioannou+25,Koehn+25}). Constraints from high mass NSs are particularly important as they help distinguish different families of EoS \citep{Bednarek+12,Han+20,Tolos+20,Legred+21,Malik+22}. Moreover, multiple radius measurements at the same mass provide the possibility to cross-check the results and tighten the EoS constraints. One radius measurement for a high mass MSP, namely \josfy, has already been performed \citep{Riley+21,Miller+21, Salmi+22,Salmi+24_J0740,Dittmann+24}, and the $\sim 2\,$\Msun MSP \jsflong (hereafter \jsf, the subject of this paper) is a prime target for radius estimation. It was also argued that, even with the low count rate from \jsf ($\sim 0.02\,$cts s$^{-1}$ with NICER, \citealt{Wolff+21}), it would be possible to obtain meaningful radius constraints given the initially observed high pulsed fraction \citep{Pancrazi+12}.

\jsf was first discovered as a $\gamma$-ray source in the third EGRET catalog \citep{Hartman+99}. It was later found to exhibit $\sim 317$ Hz pulsations using the Parkes radio telescope \citep{Crawford+06}. Radio observations with longer baseline resulted in the detection of the Shapiro delay, allowing measurement of the mass, which was found to be close to 2 \Msun \citep{Demorest+10}. Later analyses, with an even longer baseline, confirmed and refined this value \citep{Shamohammadi+23,NANOGrav_15yrs}. The tightest current estimate with a twelve year baseline is $1.937\pm0.014$ \Msun\footnote{The credible intervals are reported as the equally tailed 68\% credible interval throughout this article.} \citep{NANOGrav_15yrs}, making \jsf the second most massive MSP confirmed with radio timing after \josfy \citep{Cromartie+20,Fonseca+21}. It has since been observed with both the X-ray Multi-Mirror Mission (XMM-Newton, XMM hereafter) and NICER, which both confirmed its detection in X-rays and the presence of pulsations at these wavelengths \citep{Pancrazi+12,Wolff+21}. Archival Chandra observations of this source are also available \citep{Roberts+07}.

In Section~\ref{sec:data}, we detail the analyzed data and the various reduction steps that were undertaken. In Section~\ref{sec:modeling} we describe PPM including updated fits to the oblateness and surface gravity relations, the Bayesian inference procedure, and the priors that were used during sampling. In Section~\ref{sec:results}, we present the results of the various inference runs that were carried out, especially for our preferred model. In Section~\ref{sec:discussion} we discuss our results before finally concluding in Section~\ref{sec:conclusion}.

\section{X-ray event data}
\label{sec:data}
We now summarize the data used during inference and the various reduction steps that have been taken. All the data products used for this analysis can be found in the Zenodo repository doi:\href{https://doi.org/10.5281/zenodo.22163155}{10.5281/zenodo.22163155} \citep{zenodo}. 

\subsection{NICER}
\label{sec:NICERdata}

% Which data
The available NICER data for this source span the 2017 July 4th -- 2023 May 16th period (OBSIDs 0060310101 to 6060310220). Apart from 3 observations in July 2019 (OBSIDs 2060310246 to 2060310248, amounting to $\sim 3.2\,$ks of exposure), for which NICER experienced a time stamp anomaly, it is comprised of all available exposures antecedent to the NICER light leak\footnote{\href{https://heasarc.gsfc.nasa.gov/docs/nicer/analysis_threads/light-leak-overview/}{https://heasarc.gsfc.nasa.gov/docs/nicer/analysis\_threads/light-leak-overview/}}. The data after the light leak have high background and are not suited for analyses of sources as faint as \jsf, which has a flux ten times lower than the background before the light leak. Moreover, data after the light leak has a cumulative raw exposure of $\sim 45.5$\,ks, which is less than $5\%$ of the total exposure time. Hence, we opted not to use the data after the light leak.

The data were processed with the \textsc{Heasoft} software version 6.36, including the NICER calibration file \texttt{xti20240206} and data analysis software \textsc{NICERDAS}v15. After calibration, data were reduced using the \texttt{nicerl2} task with standard filtering options, which resulted in $\sim 1.137$ Ms of calibrated exposure. We further selected good time intervals (GTIs) using \texttt{psrpipe} to remove times of high background. The criteria to select GTIs broadly follow the procedure from \cite{Bogdanov+19_I}, and are similar to previous analyses. The exact criteria are provided in the Zenodo repository, in the form of processing scripts \citep{zenodo}.

The resulting exposure time was primarily affected by two filtering criteria: the maximum median undershoot rate and the 2–10 keV count-rate threshold. Undershoots correspond to detector resets and are correlated with an enhanced low energy background. We therefore required the median undershoot rate to be $\le200$\, cts/s. In addition, the 2–10 keV count rate provides a useful proxy for the background, as only a negligible number of source counts are expected in this energy band ($\sim$0.2\% of all counts). We thus retained only GTIs for which this count rate was below 1.5 cts/s. These values were selected to keep a long exposure time while limiting high background periods, as the source is faint and the background will be indirectly constrained by XMM observations of the target (see Section~\ref{subsec:background}). This procedure resulted in a final dataset with an exposure of $\sim946$\,ks and 1,240,139 events out of the initial $\sim7$ million events. Finally, the associated ancillary response file (ARF) and response matrix file (RMF) were extracted using the \texttt{nicerl3-spect} task. 

These events were then phase-folded using the \texttt{photonphase} task from the PINT software \citep{luo_pint_2021}. For this, we used a timing solution incorporating 11,173 times of arrival from both Nançay Radio Telescope and Fermi, spanning the entire NICER observation window. The resulting bolometric pulse profile is composed of a single broad peak, which is consistent with the one obtained by \cite{Wolff+21}. 

\begin{figure}[t]
    \centering
    \includegraphics[width=\linewidth]{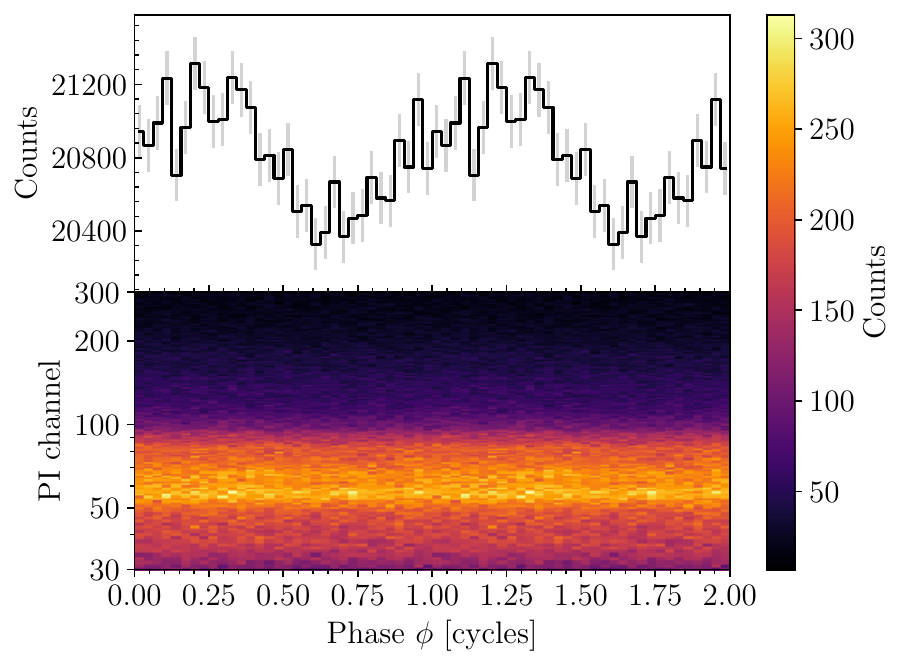}
    \caption{Data for \jsf obtained after reduction and phase-folding of the NICER events. The top panel displays the $0.3-3.0\,$keV pulse profile, displayed with the associated Poisson error in gray. The bottom panel shows the energy-resolved pulse profile of \jsf. Phases are divided into 32 equal-size bins per cycle and the energies are split by PI channels. The plotted data are the exact same data that are used for inference. The data are duplicated over two rotational phases for visualization purposes.}
    \label{fig:NICERdata}
\end{figure}

For our analysis, we used the NICER events with pulse invariant (PI) channels [30,300] which correspond to the $0.3-3.0\,$keV energy range. The significance of the pulsations in this $0.3-3.0\,$keV range ($\sim 8.5\sigma$ using the H-test, \citealt{deJager+10}) and the signal to noise ratio (SNR, $\sim0.05$) are lower than for all of the previously analyzed sources. While \josfy had only slightly better SNR, its pulse profile features two distinct sharp pulses, whereas \jsf has only one broad and skewed pulse (see Figure~\ref{fig:NICERdata}). A Kolmogorov-Smirnov test comparing the phase distribution of photons with energy above $3\,$keV to the uniform distribution found no significant presence of pulsed emission ($p \simeq 0.29$).

\subsection{XMM}
\label{sec:XMMdata}
The archival XMM data for this source are composed of three independent exposures (OBSIDs 0304960101, 0404790101 and 0655020101). With XMM, we are interested in the phase-averaged source spectrum (see Section \ref{subsec:background}). Although XMM has a timing resolution of $\sim 0.03\,$ms in Timing mode, we barely detected pulsations at a significance of $\sim3\sigma$\footnote{The \cite{Pancrazi+12} detection using the same observation, but older calibration, was at $4\sigma$ significance.}. Hence, we only used XMM data taken in imaging mode, which led us to exclude the PN data from OBSID 0655020101. 

The data were processed with the XMM Science Analysis Software \textsc{SAS} version \texttt{22.1.0} and the calibration file version \texttt{XMM-CCF-REL-417} \citep{XMM_SAS}. After calibration, we removed soft proton flares and chose events with the recommended filters (FLAG $=0$ and PATTERN $\le4$ for PN and PATTERN $\le12$ for MOS1/2). For all the observations, we used the same source region across observations and the same background region to extract the spectra. We computed the scaling of each spectrum, ARF, and RMF with, respectively, the \texttt{backscale}, \texttt{arfgen}, and \texttt{rmfgen} tasks.

The source spectrum of \jsf was extracted from a $45''$ radius circular region, where we excluded a $25''$ circular region around a nearby source located $36''$ away in order to avoid potential contamination of the spectrum. This results in a reduction by $\sim20\%$ of the area of the source extraction region. Given their respective flux in the $0.2-4.5\,$keV band and the point spread function of the EPIC instruments, we estimate that the neighbor source accounts for $\sim10\%$ of the spectrum in this band before exclusion, and $\lesssim2\%$ after. Moreover, we note that this contaminating source has a hard power law-like spectrum, which could be linked to the previously detected power law component of \jsf (see Appendix \ref{app:powerlaw}). The background was extracted from the same charge-coupled device chip as the source, in a $60''-280''$ annulus region centered on \jsf. To avoid contamination of the background spectra by nearby sources, we excluded regions $\le 25''$ around sources detected in the 4XMM-DR14 catalog with a maximum likelihood $\ge 10$ \citep{XMMcatalog_IRAPserver, XMMcatalog_VIII, XMMcatalog_IX, XMMcatalog_X}.

\begin{figure*}
    \centering
    \includegraphics[width=\linewidth]{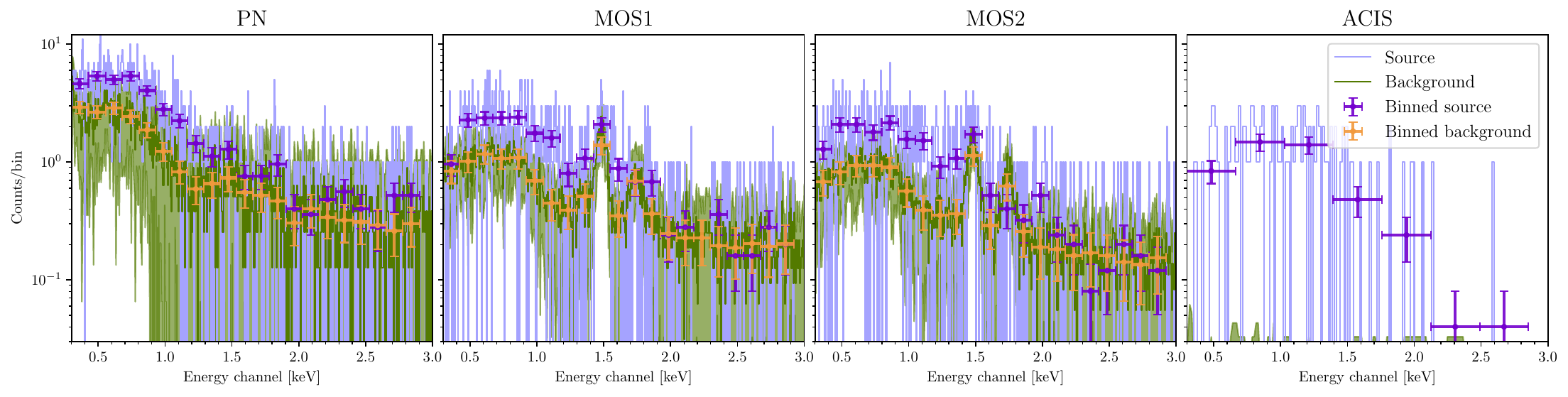}
    \caption{Count spectra of the EPIC and Chandra instruments used in the present work. From left to right, PN, MOS1, MOS2 and ACIS. The source spectra are in blue, and the rescaled background spectra in green. The green shaded region represents the 3$\sigma$ uncertainties on the background spectra. For visualization purposes only, the binned source and background spectra are provided with the associated 1$\sigma$ error bars in violet and orange, respectively. For ACIS, the background is barely visible.}
    \label{fig:EPIC_Chandra_data}
\end{figure*}

As the XMM spectra are crucial to constrain the source spectrum (see Section \ref{subsec:background}), we want to maximize the exposure time. However, with the count statistics of individual observations being low (a few tens to a few hundred source counts per observation in the energy band of interest, dominated by background), we chose to combine the different exposures into one individual source and background spectrum per EPIC instrument using \texttt{epicspeccombine}. We use the resulting exposures given in Table \ref{tab:XMMdata} to scale the source and background spectra. For our analyses, we used energy channels [56,600] for PN and [60,600] for MOS1/2, which all correspond to the $0.3-3.0\,$keV energy range. This reduced XMM data is shown in Figure \ref{fig:EPIC_Chandra_data} alongside the Chandra data.

\begin{table}[t]
\caption{Exposures in kiloseconds of the final \jsf spectra with the XMM EPIC and Chandra instruments.}
\begin{ruledtabular}
\begin{tabular}{ l c c c c c }
OBSID & Date & PN & MOS1 & MOS2 & ACIS \\
\hline
0304960101 & 2005-08-17 & 4.1 & 5.7 & 3.7 & N/A \\
0404790101 & 2007-02-08 & 24.2 & 40.8 & 39.9 & N/A \\
0655020101 & 2011-02-12 & N/A & 20.5 & 20.9 & N/A \\
\hline
7509       & 2007-04-26 & N/A & N/A & N/A & 19.8\\
\hline 
Total      &         & 28.3 & 66.8 & 64.5 & 19.8\\
\end{tabular}
\end{ruledtabular}
\label{tab:XMMdata}
\end{table}

\subsection{Chandra}
\label{sec:ChandraData}

An additional 19.8 ks Chandra/ACIS-S exposure of \jsf taken on 2007 April 26$^{\rm th}$ was also available (OBSID 7509). These observations do not have a timing resolution sufficient to observe pulses of MSPs, and can only produce phase-averaged spectra. Chandra data have already been used in PPM analysis of both MSPs \citep{Gonzalez-Caniulef+26} and quiescent low mass X-ray binaries \citep{Kazantsev+26}.  However, we still test the effects of using it alongside XMM for inference (see Section \ref{subsec:different_data}), as it has been found to exhibit $\sim 10\%$ differences in flux calibrations compared to XMM \citep{Nevalainen+10,Schellenberger+15,Marshall+21}.

The Chandra spectrum is valuable because of its low background. Moreover, its higher spatial resolution allows us to fully exclude the nearby source that slightly contaminates the XMM spectra. However, its lower exposure time and effective area make the total number of counts from the source much lower (113 cts), making it only weakly informative. This low number of counts over $\sim20$ ks implies the detection of one count every 175 seconds on average, which makes the risk of pile-up nonexistent for this source.

The Chandra data were processed using the \texttt{chandra\_repro} task from CIAO 4.18.0 \citep{Fruscione+2006, Fruscione+2026}. The source and background spectra were extracted using the \texttt{specextract} task, with the option \texttt{energy=0.25:11.0:0.01} (see Appendix \ref{app:Chandra} for more details about this option). Their respective extraction regions were chosen as a circle centered on \jsf of radius $2''$, and an annulus of inner and outer radii of $5''$ and $28''$, centered on the source as well. This choice of background region avoids all the nearby sources that could be present in the background. This procedure also produces the associated ARF and RMF files that were used for the PPM analysis. To be consistent with the NICER and XMM energy range, we chose to fit the Chandra channels [21,206] corresponding to the same $0.3-3.0$\,keV energy range.

\section{Modeling}
\label{sec:modeling}
We model the X-ray pulsations of \jsf with the X-Ray Pulse Simulation and Inference software (X-PSI, \citealt{xpsi}). In this section, we describe the modeling, including the updated fits of the oblateness and surface gravity relations, and the priors used throughout our analysis.

\subsection{Ray tracing}

In X-PSI, a surface geometry and a uniform temperature are defined for each hot spot component, with 4 to 8 parameters per spot depending on the complexity of the geometry. These hot spot geometries are considered to exist on the surface of an oblate NS, and are discretized into equally sized square cells. In this work, the hot spots are discretized in $16 \times 16$ individually emitting cells. Using the Oblate-Schwarzschild approximation (see for example \citealt{Morsink+07}), we then perform ray tracing with a special relativistic correction over 100 emergent rays from each cell, in order to compute the spectrum at a given rotational phase. This operation is performed over 64 energy values, and 32 rotational phases. The parameter values mentioned here correspond to the low resolution (LR) X-PSI settings of \cite{Mauviard+25}, which were adopted for all the runs of the present work. The specific intensity emitted by each grid cell is taken from a pre-computed NSX atmosphere table \citep{Ho+01}, as in all previous X-PSI studies of MSPs. 

Previous studies of the binary system showed that it likely underwent Roche lobe overflow while the donor star was still on the main sequence \citep{Tauris+11,Tauris+12}. Under this assumption, the accreted material was likely hydrogen-rich, which would imply a hydrogen atmosphere atop the NS as other heavier element would sink on $\sim$ second timescales \citep{Bildsten92}. The helium and carbon atmosphere cases cannot be overlooked due to possible diffuse hydrogen burning \citep{Chang+03,Chang+04,Chang+10,Wijngaarden+20}. However, a carbon atmosphere is unlikely as a simple preliminary run showed that it yielded a much lower evidence than the hydrogen case. Hence, we chose to test runs with fully ionized hydrogen, partially ionized hydrogen, and fully ionized helium atmosphere NSX models (see Section \ref{subsec:atmosphere_fit}).

\subsection{Universal Relations}
\label{subsec:QUR}

The oblateness and surface gravity of the rapidly rotating NS are necessary ingredients used for ray tracing in X-PSI. This relies on quasi-universal relations (QUR) to avoid re-computing the MSP equilibrium, which is orders of magnitude slower than the ray-tracing. In practice, these relations are almost EoS independent and have as input parameters the compactness in natural units $C=M/$\Req (with \Req the equatorial radius), the dimensionless angular velocity $\bar{\Omega}$, and the colatitude $\theta$ \citep{AGM14}. The oblateness and surface gravity relations used in X-PSI are, respectively, Equations~20 and 50 in \citealt{AGM14}. However, these were obtained using only seven distinct EoS that may not be representative. Here, we reevaluate these relations over a larger set of EoS, with a mass range and frequency value fixed to that of \jsf \citep{NANOGrav_15yrs}. 

In practice, we randomly select 1000 EoS from the publicly available tool CUTER \citep{Davis+24}\footnote{The tool is available on their associated Zenodo repository doi:\href{https://doi.org/10.5281/zenodo.10781539}{10.5281/zenodo.10781539} \citep{Davis+24_Zenodo}.}. The semi-agnostic constructions are based on the meta-model at low density (up to the nuclear saturation density) constrained by chiral effective field theory, and piecewise polytropes at higher density. It also guarantees a consistent calculation of the neutron star crust.

For each EoS, we compute the equilibrium of a rotating NS, assuming the rotational frequency of \jsf, using using a code directly adapted from the \texttt{LORENE/nrotstar}\footnote{\href{https://gitlab.in2p3.fr/lorene/Lorene}{https://gitlab.in2p3.fr/lorene/Lorene}} numerical relativity tool \citep{Bonazzola+93,Gourgoulhon+01,Gourgoulhon+10}. This allows us to retrieve the surface gravity and distance of the surface to the center of mass over 17 latitude points, and 10 central enthalpy values associated with masses within $\pm5\sigma$ of the measured mass of \jsf (see Section~\ref{sec:introduction}, and \citealt{NANOGrav_15yrs}). With these, it is then possible to find fits to the QURs tailored for \jsf, skipping the interpolation over a frequency grid, as was done before. This also makes the evaluation of the fit faster, as only one frequency needs to be computed.

Using PySR \citep{PySR}, we then performed symbolic regression on the data points to find functional forms that best represent them. For the surface gravity, we extracted a functional form of $g(\theta)/g_0$, where $g(\theta)$ is the surface gravity at colatitude $\theta$ and $g_0$ is the surface gravity for a spherical star of same rotating mass and equatorial radius. For the oblateness, using the same procedure, we found a functional form of the dimensionless radius ratio $\bar{\mathcal{R}} = (R_{\rm polar} - R_{\rm eq})/$\Req, where $R_{\rm polar}$ is the polar radius. Subsequently applying the same $R(\theta)=R_{\rm eq} ( 1 + \bar{\mathcal{R}}\cos^2(\theta))$ expansion as used in \cite{AGM14} provides the radius as a function of colatitude. Both functions depend on the compactness and dimensionless angular velocity and are of the following forms, whose parameter values can be found in Table \ref{tab:QUR}:

\begin{equation}
    \begin{split}
        \frac{g(\theta)}{g_0} & = b + \bar{\Omega}^2 [ b_0 + \left(b_{20} + b_{21} C\right)\cos^2(\theta) + b_{40} C \cos^4(\theta) ], \\
        \bar{\mathcal{R}} & = \bar{\Omega}^2 (a_0 + a_1 C + \frac{a_2}{C^2} + \frac{a_3}{C^4}).
        \label{eq:gnrom_and_rbar}
    \end{split}
\end{equation}

\begin{table*}
    % \centering
    \caption{Values of the parameters in the functionals obtained for the two QUR (Equation \ref{eq:gnrom_and_rbar}). The associated values of \cite{AGM14} are also provided, although the reader must keep in mind that they used a different functional form so they are not directly comparable.}
    \label{tab:QUR}
    \hskip-2.5cm \begin{tabular}{c|cccc|ccccc}
            \hline
            \hline
             Reference &  $a_0$&  $a_1$&  $a_2$&  $a_3$&  $b$&  $b_0$&  $b_{20}$& $b_{21}$ & $b_{40}$\\
             \hline
             This work &  0.615&  -2.0&  -0.0539&  0.00079&  0.9985&  -0.5454&  1.504&  0.2422& -0.6679\\
              \cite{AGM14} & -0.788&  1.030&  0&  0&  1.0&  -0.791&  1.929&  -2.207& 0\\
              \hline
              Equivalent parameter in & $a_0$ or $o_{20}$& $a_1$ or $o_{21}$& N/A& N/A& Fixed & $c_{0,e}$& $c_{0,p} - c_{0,e}$& $c_{1,p} - c_{1,e}$&N/A\\
              \cite{AGM14} &&&&&&&&&\\
            \hline
        \end{tabular}
    \end{table*}

\begin{figure}[t]
    \centering
    \includegraphics[width=\linewidth]{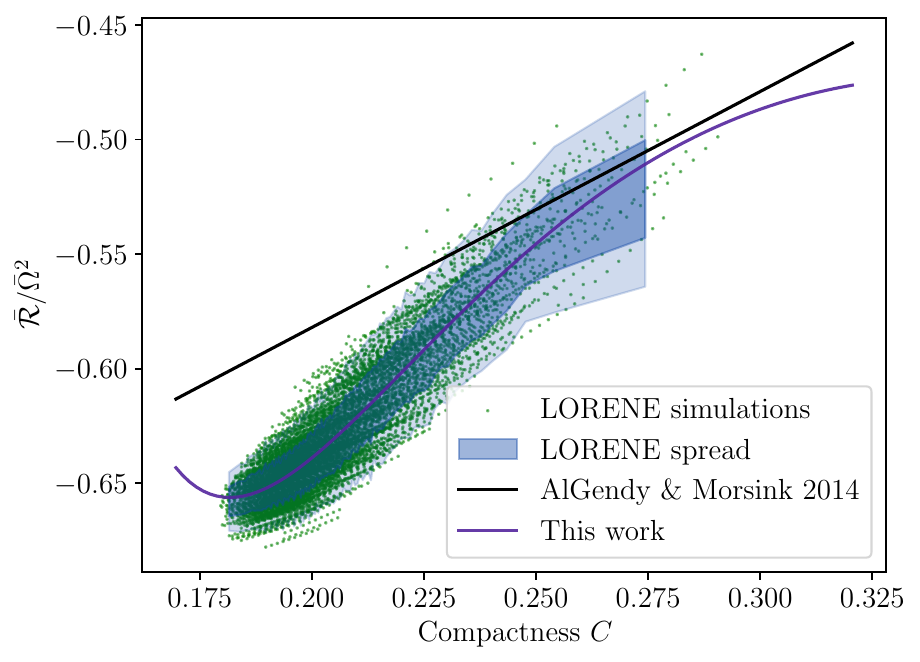}
    \caption{Scatter plot of the dimensionless radius ratio $\bar{\mathcal{R}}/\bar{\Omega}^2$ with respect to the compactness, computed using the adapted \texttt{LORENE/nrotstar} tool for 1000 EoS with 17 latitude points. For each EoS, we compute $\bar{\mathcal{R}}$ for 10 central enthalpy values, chosen so that the masses are all within $\pm5\sigma$ of the mass range of \jsf. The QUR from \cite{AGM14} is shown in black, while the reevaluated functional for \jsf is in purple. The statistical distribution (68\% and 95\%) of the points is represented in blue.}
    \label{fig:QUR}
\end{figure}

The $\bar{\mathcal{R}}$ values computed with the adapted \texttt{LORENE/nrotstar} tool are shown in Figure \ref{fig:QUR}, alongside the fit from \cite{AGM14} and our updated functional. We stress that our functional is specific to the frequency and mass of \jsf, while the previously used formula was valid across a large mass and frequency range. Our updated functional agrees with the previously used general formula at higher compactness (for $C \gtrsim 0.24$), where the polar flattening is between one and two percent in both cases. There is, however, a $\sim 10\%$ difference in $\bar{\mathcal{R}} / \bar{\Omega}^2$ at lower compactness. This corresponds to an enhanced oblate shape, with polar flattening increased from 3\% to 4\% for our updated functionals tailored to the spin frequency and mass of \jsf. To investigate whether using these updated fits of the QUR would affect the inferred compactness, we compared results from X-PSI inferences with both the general and the updated functionals. This showed that both provide consistent results, as the data from this source has a tendency to favor higher compactness (see Section~\ref{subsec:testQUR} for more details). 

The residuals of the functionals from \cite{AGM14} to the data points from the adapted \texttt{LORENE/nrotstar} tool are asymmetric and $\sim10\%$ for both quantities, reaching up to $\sim 15\%$ for the oblateness relation. Using the new functional, we find normal distributed centered residuals with $0.03\%$ and $2\%$ standard deviation respectively for the effective surface gravity and oblateness. The associated residual plots and effective gravity data plot are available in the Zenodo repository \citep{zenodo}.

\subsection{Instrument}
\label{subsec:instrument}

The modeled spectrum is convolved with the response of the instruments (RMF $\times$ ARF) to determine the signal as it would be seen by the instrument and to evaluate the likelihood when comparing the model to the data. To account for cross-calibration uncertainties between the different instruments, we multiply each ARF by a free energy-independent scaling factor parameter.

Similarly to previous X-PSI analyses, we assumed Gaussian priors on $\alpha_{\rm NICER}$, $\alpha_{\rm EPN}$, $\alpha_{\rm EMOS1}$, and $\alpha_{\rm EMOS2}$. We had intended to do the same for ACIS. Unfortunately due to a typo in the scripts we assumed a Uniform prior instead: $\alpha_{\rm ACIS}\sim\mathcal{U}(0.5,1.5)$, making the prior more conservative. Importance sampling using our intended prior determined that this typo made no difference to the final result. Hence, we chose not to re-run all inferences and keep this slightly more conservative result. As the knowledge of the absolute ARF values of individual instruments is better than the cross-calibration between different instruments, we tried one run where we fixed $\alpha_{\rm EPN}$ and tightened the prior of the MOS1/2 energy-independent scaling factor down to $\alpha \sim \mathcal{N}(\mu=1.0, \sigma=0.03)$. This resulted in similar radius and geometric parameters constraints as using the default cross-calibration prior, hence we opted to use the more conservative default prior. 

Unlike previous X-PSI analyses, instrument responses are now trimmed automatically. For each channel selected for fitting, we compute the normalized cumulative of the redistribution over increasing incident photon energies. We reject the energies where the complementary normalized cumulative of all the channels are smaller than a lower quantile, defined by the user as a tolerance. For this analysis, we set the tolerance parameters to be $10^{-4}$ ($0.01\%$) for all the instruments, meaning that we keep $\ge99.99\%$ of the effective surface from the response. This procedure allows us to keep most of the response while avoiding computing the model to unnecessarily high energies. 

\subsection{Background}
\label{subsec:background}

The background is considered to be phase-independent, as no process is expected to occur at the exact same 317 Hz spin frequency of \jsf. For NICER data, contrary to XMM and Chandra, it is not possible to empirically estimate a background component as the instrument does not have imaging capabilities. To circumvent this issue, using parametric or empirical NICER background models, such as SCORPEON\footnote{\href{https://heasarc.gsfc.nasa.gov/docs/nicer/analysis_threads/scorpeon-overview/}{https://heasarc.gsfc.nasa.gov/docs/nicer/analysis\_threads/scorpeon-overview/}} \citep{Markwardt+24} or 3C50 \citep{Remillard+22} is possible \citep{Choudhury+24_J0437,Miller+26}. Another method that was previously used, is to jointly fit the NICER data with a spectro-imaging instrument, like XMM-EPIC. This ensures that the inferred NICER source spectrum is consistent with the observed XMM source spectrum \citep{Riley+21,Miller+21}. This method is not background model dependent and previously proved to indirectly constrain well the NICER background \citep{Salmi+22,Salmi+24_J0740,Salmi+24_J1231,Vinciguerra+24_J0030,Mauviard+25,Kini+26}. We chose here to use the latter method with both XMM and Chandra data to avoid dependency on the background model.

In practice, the XMM and Chandra background spectra are left free to vary within $\pm4\sigma$ of their measured values, while the NICER background is left unconstrained. To estimate the likelihood, we marginalize over the phase-independent background for all instruments (see Appendix B in \citealt{Riley19_PhD} for more details on this procedure).

\subsection{Priors}
\label{subsec:prior}

For the mass, inclination and distance priors, we use the values from the 15 year dataset of the NANOGrav collaboration \citep{NANOGrav_15yrs}. The mass and inclination values follow Gaussian distributions with $M = 1.937\pm0.014$\Msun and $\sin(i) = 0.9999019\pm0.0000037$. Using the same radio timing solution, the distance prior is derived from the radio parallax measurement to be $D=678\pm42$\,pc\footnote{The parallax value is extracted from the timing solution files provided \citep{NANOGrav_15yrs_Zenodo}.}.

For the radius, we use a broad uniform prior spanning $6-16$\,km. In practice, however, we reject samples that go above a compactness limit defined as $R_{\rm polar}/r_g(M) > 3$, with $r_g(M)=GM/c^2$ the gravitational radius for a star of mass $M$. We also reject samples reaching effective surface gravity outside of atmosphere tables bounds (for all colatitudes $\theta$, $\log_{10}(g(\theta)\,\rm{[cm/s^2]}) \le 15.0$ for the fully ionized hydrogen model). The atmosphere table upper effective surface gravity bound implies a minimum value of the radius, which is $R_{\rm min} \simeq 8.5\,$km for \jsf given its mass. There is no effect from these limits on the upper radius boundary which remains 16 km.

\cite{Pancrazi+12} estimated the hydrogen column density parameter \nh to be $\sim (1-4) \times 10^{21}\,\rm{cm}^{-2}$ from X-ray spectral fitting. The \cite{hi4pi_2016} also found the galaxy integrated \nh to be $\sim 10^{21}\,\rm{cm}^{-2}$ in the direction of the pulsar. Hence, we use a broad uniform prior on the \nh spanning $10^{17} - 10^{22}\,\rm{cm}^{-2}$. Test runs using a log-uniform prior over the same range yielded similar results.

The geometric priors are mostly the same as in previous analyses, however, the uniform temperature priors previously spanning the bounds of the NSX table have changed. These are more restricted than before, now spanning $\log_{10}(T\,[\rm K]) \sim \mathcal{U}(5.5, 6.5)$ instead of the $(5.1, 6.8)$ range used in previous X-PSI analyses (see next section).

\subsection{Degeneracies for low temperatures}
\label{subsec:degeneracies_T}

The new $\log_{10}(T\,[\rm K]) \sim \mathcal{U}(5.5, 6.5)$ range for the temperature prior has been carefully selected. While the upper limit is chosen to be the same for all NSX tables\footnote{We also note that no previous analysis of NICER data yielded temperature above $\log_{10}(T\,[\rm K]) \sim 6.5$.}, so that their evidences are easily comparable (i.e., same prior volumes), the lower limit has been selected to reduce degeneracies. These arise because \jsf is faint and its bolometric pulse profile does not exhibit clear departure from a sinusoid, likely due to its low source count statistics (see Figure \ref{fig:NICERdata}). Hence, a single hot spot fits the data, although noticeably worse than two hot spots (log-likelihood is lower than the two hot spots case by $\sim 10$). Due to large interstellar absorption ($\sim 10^{21}\,\rm{cm}^{-2}$) and low count statistics, spots with temperature up to $\log_{10}(T[\rm K]) \sim 5.7$ can remain undetectable in our data set, which resulted in a large part of the parameter space being unconstrained when using a two hot spot model. Such a phenomenon did not arise for \josfy, which had two distinct pulses that strongly favored the presence of two hot spots on its surface.

For \jsf, the likelihood surface can be pictured as a large flat region with acceptable likelihoods, representing the case where only one hot spot is detectable. On top of this flat region, there are local peaks (increase in log-likelihood of $\sim10$) corresponding to the case where both hot spots are visible. These local peaks are identified as individual modes by MultiNest, while the flat region is outside of these modes. Using $\log_{10}(T[\rm K]) = 5.5$ as a lower limit for our temperature prior allows us to retain the full distribution of the individual modes, while increasing the local log-evidence of these individual modes by $\sim 3$. This change is much greater than what is expected from a uniform prior change (that we estimate to be $\sim 0.5$), and hints that the weight of the individual modes were likely underestimated.

Temperatures of hot spots down to $\log_{10}(T\,[\rm K]) \sim 5.1$ are not expected, and such a temperature would imply that a hydrogen atmosphere would be only partially ionized at equilibrium (up to a few percent of neutral hydrogen), which is incompatible with the fully ionized atmosphere model we use. Moreover, no previous NICER analysis resulted in a temperature of the hot spots being lower than $\log_{10}(T\,[\rm K]) \sim 5.5$. Finally, the previous choice of lower limit was not physically motivated, and was made to use the whole range of the atmosphere tables. Hence, this new lower limit on temperature has been adopted for all the following runs.

For the \STU geometry (see Section~\ref{subsec:ST-U}), we further reduce degeneracies by imposing an ordering on the temperature of the two individual ST hot spots. The primary ST hot spot is now required to be hotter than the secondary one, whereas the primary/secondary ordering was previously based on colatitude. This new choice forces the secondary hot spot temperature to be the single partially unconstrained parameter, rather than leaving both the primary and secondary temperatures partially unconstrained.

\subsection{Sampling}
\label{subsec:sampling}

We perform nested sampling with MultiNest \citep{MultiNest_2009,PyMultiNest}. Parameter recovery for PPM has been shown to be reliable using MultiNest given appropriate sampler settings, and has been tested extensively \citep{Vinciguerra+24_XPSI}. MultiNest has been benchmarked against the more robust UltraNest \citep{buchner_ultranest_2021} for the inference case of \josfy, a source with similar SNR, mass, and inclination priors as \jsf. This yielded consistent results for both samplers, with UltraNest being $\sim 10$ times slower than MultiNest \citep{Hoogkamer+25}. Due to UltraNest being much more computationally expensive, we therefore use MultiNest for our exploratory and headline runs.

Based on preliminary tests, we settled on using $2\times10^4$ live points (LP), a sampling efficiency (SE) of $0.05$ and evidence tolerance (ET) termination criterion of $0.1$ for the preliminary runs. Convergence is also faster for \jsf, as fewer likelihood estimations are required due to the lower SNR. Hence, we were able to use these improved sampler settings compared to most previous NICER analyses, which allowed for a thorough sampling of the likelihood surface, even for exploratory runs. For the headline run, we further improved these to $4\times10^4$ LP and $0.03$ SE, while keeping the ET fixed. With these settings, the computation time was  $\gtrsim 10$ longer and our available computational resources did not allow us to increase LP or reduce SE further. However, these settings are on a par with the highest resolution ever used for PPM \citep{Salmi+24_J0740} and the results appear to be converged.

Moreover, we use the multi-modal version of MultiNest, which should enable faster computations for multi-modal likelihood surfaces. This method uses a clustering algorithm to split modes, which is performed by default using the all parameters. However, MultiNest also permits to define a subset of parameters for mode splitting, but does not easily allow to individually pick them. We know that some parameters, such as as the cross-calibration factors, show only minor, or nonexistent, multi-modal features. However some parameters are highly multi-modal, such as the parameters defining the geometry and temperature of the hot spots. Hence to improve the mode splitting we chose to perform the split based on the mass, radius, distance, inclination and hot spot parameters. 

\section{Results}
\label{sec:results}

In this section, we first investigate the effects of the new QUR implementation and the results for a variety of data sets, surface geometric models and atmosphere tables. Then, we describe the result from a higher resolution headline model that is recommended for downstream EoS inference. Results and posterior distributions for all of the investigated models are provided in the Zenodo repository \citep{zenodo}.

\subsection{Updated QUR fits}
\label{subsec:testQUR}
We assessed the effect on PPM of using the new QUR fits tailored to the rotational frequency of \jsf (see Section~\ref{subsec:QUR}). This was done by comparing the inference results using the previous QUR fits from \cite{AGM14} to our updated functionals. These runs both include a fully ionized hydrogen atmosphere with a simple \textit{Single Temperature Unshared} (\STU) geometry, with two circular, independent and uniformly emitting hot spots, individually referred to as ST spots (see Figure~1  of \citealt{Vinciguerra+24_XPSI} for an explanation of the hot spot naming convention).

\begin{figure}[t]
    \centering
    \includegraphics[width=\linewidth]{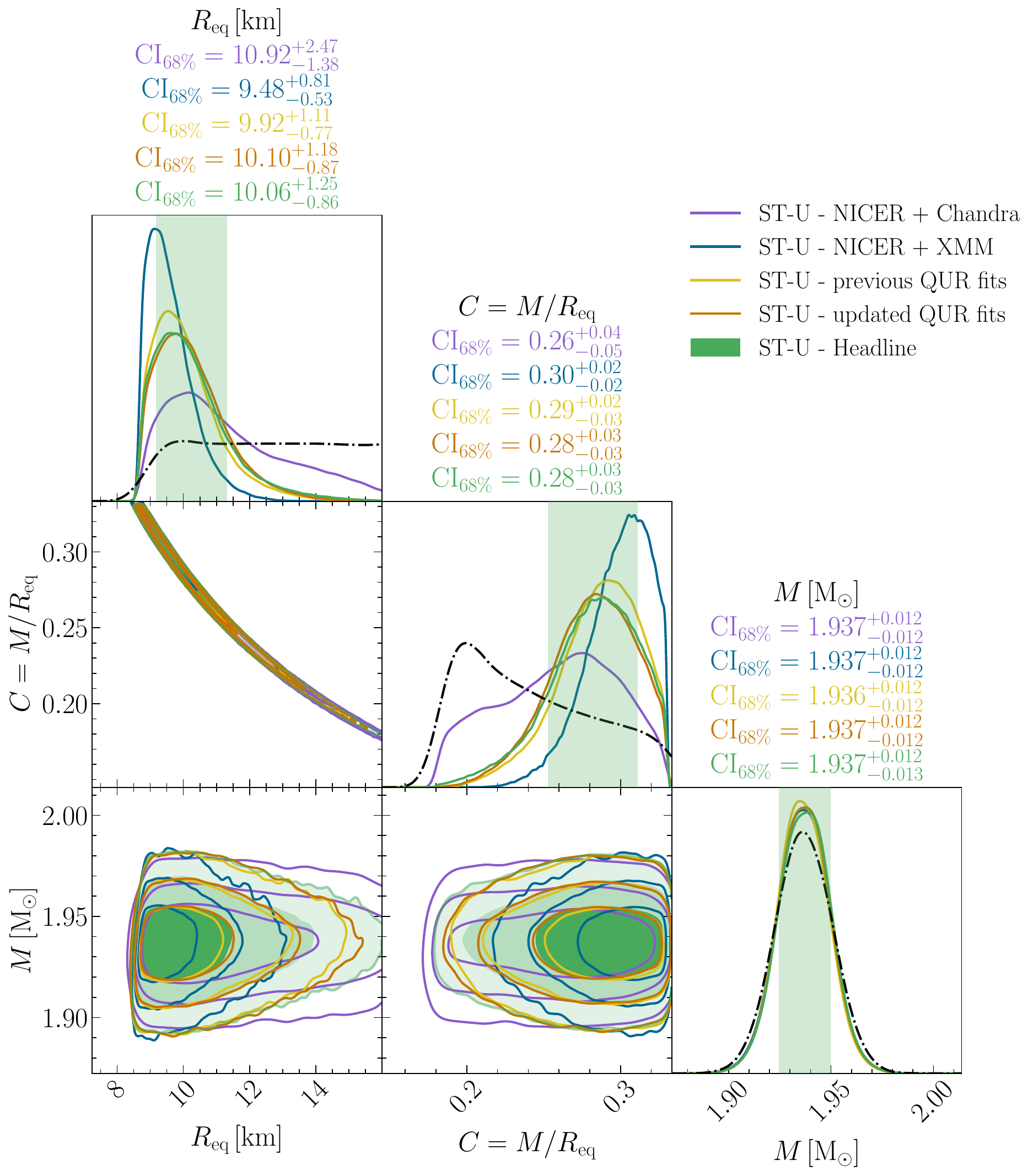}
    \caption{Posterior distributions of the radius, compactness, and mass for the \STU model, using the previous and updated fit to the QUR, different data sets, as well as the headline result. Dashed–dotted black lines on the diagonal plots show the marginalized prior distributions, shared across all runs beside the previous QUR fit one. However, if present, the prior of the previous QUR run would be hardly distinguishable. The shaded vertical bands show the 68.3\% credible intervals (CI) of the headline result. The contours in the 2D posteriors show the 68.3\%, 95.4\%, and 99.7\% credible regions, which are filled for the headline result.}
    \label{fig:MR_QUR}
\end{figure}

Inference using the updated QUR fits leads to similar model evidence and posterior distributions for all parameters. For the radius, we notice a slight extension of the high radius tail, which increases the median radius by $\sim 180\,$m and widens the 68\% CI by $\sim 180\,$m (see Figure \ref{fig:MR_QUR}). This effect is likely due to the few percent increase in oblateness introduced by our updated QUR fits (see Figure \ref{fig:QUR}), such that light rays emitted near the poles of the star require less bending to reach the observer for sources that are seen edge-on. \jsf has such a viewing geometry with a hot spot near one pole, hence, compactness does not need to be as important as it was before to explain the data, such that the radius can take higher values.

Given the small change in inferred radius, which is expected, and the increase in radius uncertainties by a comparable amount, we conclude that using the updated fits to the QUR gives results that are consistent with previously used relations \citep{Morsink+07,AGM14}. As we consider this updated fits to the QUR better representative of reality, we use it for all the subsequent runs in this study.

\subsection{Different data sets}
\label{subsec:different_data}

To test the effect of using Chandra data alongside both XMM and NICER (see Section~\ref{sec:ChandraData}), we ran inferences with the \STU geometry, using the new QUR implementation described in Section \ref{subsec:QUR}. We tested one reference dataset with NICER+XMM, one with NICER+Chandra, and one with NICER+XMM+Chandra.

For every instrument in every run, we consider possible energy-independent cross-calibration factors, that are included as model parameters and fitted (see Section \ref{subsec:instrument}). For the NICER+XMM+Chandra run, we find $\alpha_{\rm ACIS}=0.92\pm0.07$\footnote{Value found when applying the Gaussian prior via importance sampling rather than the uniform one, see Section \ref{subsec:instrument}.}, while $\alpha_{\rm EPN,EMOS1,EMOS2}\simeq1.03\pm0.07$. This is consistent with the previously found value of $\sim 10\%$ difference in effective area calibration between Chandra-ACIS and XMM-EPIC instruments in the $0.5-2.0\,$keV band, making the observed flux by Chandra lower than that of XMM \citep{Nevalainen+10}.

The inferred parameters change slightly when using different data, which is expected given the cross-calibration uncertainties. In particular, the hotter primary spot has a larger size, but with lower temperature, when using XMM instead of Chandra. The median primary temperature difference is $\Delta\log_{10}(T_p \rm{[K]})=-0.113$, which corresponds to a temperature decrease of $\sim23\%$. Differences in flux calibration have been shown to result in $23\%$ lower temperature for EPIC instruments than with Chandra for galaxy clusters \citep{Schellenberger+15}, which is consistent with the difference observed here.

The radius with NICER+Chandra data is not as well constrained as with NICER+XMM, and it goes to higher values (see Figure \ref{fig:MR_QUR}). This larger uncertainty is partially caused by the scarcity of photons in the Chandra spectrum (113 counts only). Moreover, the lower inferred flux from Chandra indirectly constrains the NICER background, and results in higher pulsed fraction. Hence, the unpulsed component is not as important making the inferred radius upper limit higher than with XMM (see Section~\ref{subsec:low_model_dependency} for a discussion on this effect).

Finally, introducing Chandra alongside NICER and XMM data only results in broadening of the posterior intervals for all parameters, all being consistent with the NICER+XMM value within $\lesssim1\sigma$. Hence, using both XMM and Chandra data together with cross-calibration factors does not result in overconfident measurements in this case. Therefore, given the value added by the additional data, we chose to use NICER+XMM+Chandra data across all subsequent runs. 

\subsection{\STU with and without additional components}
\label{subsec:ST-U}

We first present the results of exploratory runs using the \STU model.  We then test the impact of adding additional components: a warm surface or power law component. We report the values of their respective Bayesian evidence and maximum likelihood in Table \ref{tab:modelComparison}.

\label{sec:diff_models_xpsi}
\begin{table}[t]
\caption{Log-evidence $\ln \mathcal{Z}$, maximum log-likelihood $\ln p( \text{d} |\bm{\theta}_{\text ML})$, and upper 90th percentile radius $R_{\rm eq}^{90}\,$ values for the different investigated models. Uncertainties on the log-evidences are all between 0.02 and 0.04. Runs testing the effects of XMM and Chandra data (see Section \ref{subsec:different_data}) are not included here as the data sets are different and their evidences and likelihoods are not comparable.}
\begin{ruledtabular}
\begin{tabular}{ l | c c | c }
     Model & $\ln \mathcal{Z}$ & $\ln p( \text{d} |\bm{\theta}_{\text ML})$ & $R_{\rm eq}^{90}\,$[km] \\
     \hline
     \CST & $-54072.34$ & -54049.89 & 12.53 \\ %\pm 0.03
     \hline
     \STU & $-54075.06$ & -54047.14 & 11.70 \\ %\pm 0.03
     \STUEL & $-54074.19$ & -54045.44 & 12.08 \\ %\pm 0.03
     \STUPL $-$ unpulsed & $-54083.42$ & -54047.45 & 10.74 \\ %\pm 0.04
     \STUPL $-$ pulsed & $-54083.53$& -54047.69 & 10.65 \\ %\pm 0.04
     \hline
     \STU $-$ Hp & $-54076.93$ & -54046.61 & 11.55 \\ %\pm 0.03
     \STU $-$ He & $-54073.65$ & -54046.73 & 11.93 \\ %\pm 0.03
     \hline
     \STCST & $-54073.56$ & -54047.27 & 11.76 \\ %\pm 0.03
     \STCDT & $-54072.32$ & -54045.53 & 10.90 \\ %\pm 0.03
     \STPST & $-54076.25$ & -54047.14 & 12.08 \\ %\pm 0.03
     \STPDT & $-54074.04$ & -54044.95 & 11.51 \\ %\pm 0.03
     \hline
     \STU $-$ headline & $-54075.12$ & -54047.01 & 11.80 \\ %\pm 0.02
\end{tabular}
\end{ruledtabular}
\label{tab:modelComparison}
\end{table}

\subsubsection{\STU}

The \STU configuration has been tested for all MSPs analyzed so far, and was shown to provide a satisfactory fit to \josfy \citep{Riley+21,Salmi+22,Salmi+24_J0740}, the only other MSP analyzed with PPM with such a low flux ($\sim 10^{-14}\, \rm{erg\,cm}^{-2}\,\rm{s}^{-1}$, \citealt{Wolff+21}). For \jsf, \STU provides again satisfactory results.

The \STU configuration yielded two modes, each having a spot near the equator and another near the pole. However, these two modes are degenerate, one with the polar hot spot at the North pole, and the other at the South pole. In both cases, the polar hot spot is hotter, and is hence the primary hot spot. This configuration is similar to what was found for \josft \citep{Mauviard+25}. For both degenerate modes, the primary polar hot spot has similar phase $\phi_p = 0.88^{+0.04}_{-0.06}$, radius $\zeta_p \simeq 0.12^{+0.05}_{-0.03}$\,rad, and temperature $\log_{10}(T_p\,\rm{[K]}) \simeq 6.12^{+0.05}_{-0.06}$. This is also valid for the secondary equatorial hot spot with phase $\phi_s = 0.55^{+0.02}_{-0.03}$, radius $\zeta_s \simeq 0.3^{+0.3}_{-0.1}$\,rad, and temperature $\log_{10}(T_p\,\rm{[K]}) \simeq 5.85^{+0.11}_{-0.13}$. The only difference between the modes is the colatitude of their polar hot spots, which are opposite with respect to the equator ($\theta_p=0.38^{+0.15}_{-0.11}\,$rad and $\theta_p=2.77^{+0.12}_{-0.17}\,$rad), with similar equatorial spot colatitude $\theta_s = 1.6^{+0.5}_{-0.5}$\,rad.

Outside of the two modes, the posterior covers a region of the parameter space where the secondary hot spot only weakly contributes to the pulse (see Section~\ref{subsec:degeneracies_T}). Samples in this region have lower primary phase of $\phi_p \sim 0.75$ while the secondary is colder ($\log_{10}(T_s\,\rm{[K]}) \lesssim 5.7$). These parameters allow the primary alone to reproduce the data, while the the secondary has negligible contribution and is barely detectable (few percent of the source counts at most). All of the posterior structures mentioned here are also seen in the headline posterior distribution in Appendix~\ref{app:full_posterior}.

To assess the use of the orbital inclination prior as a proxy for the viewing angle, we performed one \STU run with a uniform prior on the inclination cosine $\cos(i)\sim\mathcal{U}(0,1)$ instead of the narrow prior from radio timing. This run resulted in two configurations: one infers inclination on the edge of the prior, similar to the radio timing value ($i\sim90\degree$) and yielding similar geometries, while the other finds lower inclination of $\sin(i)\sim0.5$ where one of the two spot that is barely detectable due to its low temperature. Moreover, the inferred radius is similar, although the constraint is tighter than the one found using the narrow inclination prior.

\subsubsection{Power law component}

We found no evidence for the presence of a power law component in the phase-averaged XMM and Chandra data (see Appendix \ref{app:powerlaw} for a spectroscopic study and discussion of this possible component). We further investigated the effect of adding a power law component to our PPM analysis with the \STU geometry (\STUPL).

We first considered an unpulsed power law component, with a uniform prior on the photon index $\Gamma\sim\mathcal{U}(1,2)$, and a log-uniform prior on the norm $\log_{10}(\rm norm \,[photons/cm^2/s/keV])\sim\mathcal{U}(-8,-2)$. This resulted in a similar hot spot configuration as for \STU alone, with a drastically decreased evidence, making this model strongly disfavored (see Table \ref{tab:modelComparison}). This is likely because there is no increase in likelihood, while the number of parameters increases by two. Moreover, the power law component has an upper 90th percentile normalization of $\le 1.5\times 10^{-6}\,$photons/keV/cm$^2$/s, which is on the lower edge of the prior, compatible with zero, and similar to the value found in our spectroscopic analysis. 

We also considered the possibility of a pulsed power law component using a sinusoidal form of the same period as \jsf spin. The power law was modulated by an oscillation of the form $1 + A\cos(2\pi(\phi-\phi_0))$, where $A$ is the amplitude of the oscillation, $\phi$ is the rotational phase, and $\phi_0$ is the phase shift. We considered an uniform prior on both the amplitude ($A\sim\mathcal{U}(0,1)$) and the phase shift ($\phi_0\sim\mathcal{U}(0,1)$, periodic), while keeping the unpulsed prior for the power law parameters. The conclusions are the same as when the power law was unpulsed: the model is strongly disfavored by evidence, and the power law component has an upper 90th percentile normalization of $\le 2.1\times 10^{-6}\,$photons/keV/cm$^2$/s. Hence, we conclude that we do not detect the presence of a power law component, pulsed or not, for \jsf.

\subsubsection{Warm surface component}
We also tried adding a warm surface component\footnote{Although MSPs are old sources, they may still be heated by internal mechanisms driven by spin-down, resulting in warm thermal emission from the bulk of the NS surface (see, e.g., \citealt{Rodriguez+26}, and references therein).} to the \STU geometry (also known as an Elsewhere component in X-PSI, \STUEL). This time, we considered the partially ionized hydrogen NSX atmosphere table for the warm surface only \citep{Ho+01}, as we expect the surface temperature to be low enough for neutral hydrogen to exist in non negligible quantities (few percent). Again, we retrieve a similar hot spot configuration, with one hot spot near the pole, and one near the equator. This model constrained the temperature of the warm surface to be $\log_{10}(T_{\rm EW}\,\rm{[K]})=5.56^{+0.06}_{-0.12}$, which is higher but still consistent with the warm surface temperature of \jofts ($\log_{10}(T\,\rm{[K]}) = 5.48 \pm 0.02$, \citealt{Gonzalez-Caniulef+19,Qi+26}), inferred from ultraviolet emission detected by the Hubble Space Telescope \citep{Kargaltsev+04,Durant+12}. We find a slightly increased radius by $\sim 200\,$m compared to the \STU case. The evidence does increase by $\Delta \ln \mathcal{Z} = 0.87 \pm 0.04$, which is not significant \citep{Kass+95}. Hence, we still consider \STU to perform better than \STUEL.

\subsection{Different geometries}
\label{subsec:geometric_configs}

We tried other geometries than \STU, allowing for a different temperature distribution, to possibly find more likely hot spot configurations to describe the data. For all the models presented here, we use the same data, priors, Multinest sampling parameters and X-PSI resolution settings (see Section~\ref{sec:modeling}).

\subsubsection{More complex models}
\label{subsubsec:more_complex_models}

After trying the simple \STU model in Section \ref{subsec:ST-U}, we then kept a simple primary ST hot spot while using more complex secondary hot spot geometries: \textit{Concentric Single Temperature} (\STCST), \textit{Concentric Double Temperature} (\STCDT), \textit{Protruding Single Temperature} (\STPST), and \textit{Protruding Double Temperature} (\STPDT). These more complex geometries used for the secondary hot spot are constructed by overlapping two circular regions. A \textit{Concentric} geometry describes two concentric overlapping regions, while \textit{Protruding} indicates that they must simply overlap. A \textit{Single Temperature} region in the complex spot nomenclature specifies that only one of the two region of the spot emits radiation, while the overlapping region blocks emission from the other, and \textit{Double Temperature} means that both regions of the spot each emit at their own independent temperature (refer to \citealt{Riley+19} and Figure~1 of \citealt{Vinciguerra+24_XPSI} for more details on the models and their nomenclature).

The \STCST run resulted in five modes, four corresponding to the two \STU modes where the CST component replaces either the polar or equatorial spot, and one with a large thin ring (radius of $\zeta = 71.2^{+2.4}_{-2.6}\,\degree$, angular width of $0.49^{+0.22}_{-0.14}\,\degree$, and temperature of $\log_{10}(T\,\rm{[K]})=6.10^{+0.03}_{-0.04}$) and a warm ($\log_{10}(T\,\rm{[K]})=5.71^{+0.11}_{-0.12}$) circular hot spot near the equator. This latter configuration has the highest local log-evidence for this model ($\ln \mathcal{Z}=-54074.12 \pm 0.03$) and pushed us to test the CST only geometry that is presented in Section \ref{subsubsec:CST}. The \STCDT geometry had four modes, two corresponding to \STU-like \STCST modes, one with a large (radius $\zeta = 51^{+26}_{-31}\,\degree$) and warm ($\log_{10}(T\,\rm{[K]})=5.66^{+0.10}_{-0.07}$) overlapped CDT component that is comparable to \STUEL, and one matching the large thin ring from \STCST. For this model, the mode similar to \STUEL provides the highest local log-evidence ($\ln \mathcal{Z}=-54072.37 \pm 0.03$)

The \STPST geometry resulted in two modes, each being similar to one of the \STU modes with one ST hot spot near the pole and one PST near the equator. Finally, the \STPDT inference run only found one mode that is reminiscent of the \STUEL-like \STCDT component.

For most of these, we retrieve improved evidence over the \STU model (see Table \ref{tab:modelComparison}). However, the improvement is at most $\Delta\ln\mathcal{Z} \simeq 2.72 \pm 0.04 $ for the \STCDT model. Such an improvement is described as strong but not decisive by the Kass and Raftery criteria \citep{Kass+95}. For PPM, it was also previously shown through simulations that the wrong model could be favored by a log-evidence of $\sim1.6$ \citep{Vinciguerra+24_XPSI}. To be conservative, and given the low count statistics in the pulse (see Figure \ref{fig:NICERdata}), we chose to keep the simpler \STU model for our reported result to be used for EoS inference.

\subsubsection{The CST only geometry}
\label{subsubsec:CST}
\begin{figure}[t]
    \centering
    \includegraphics[width=0.6\linewidth]{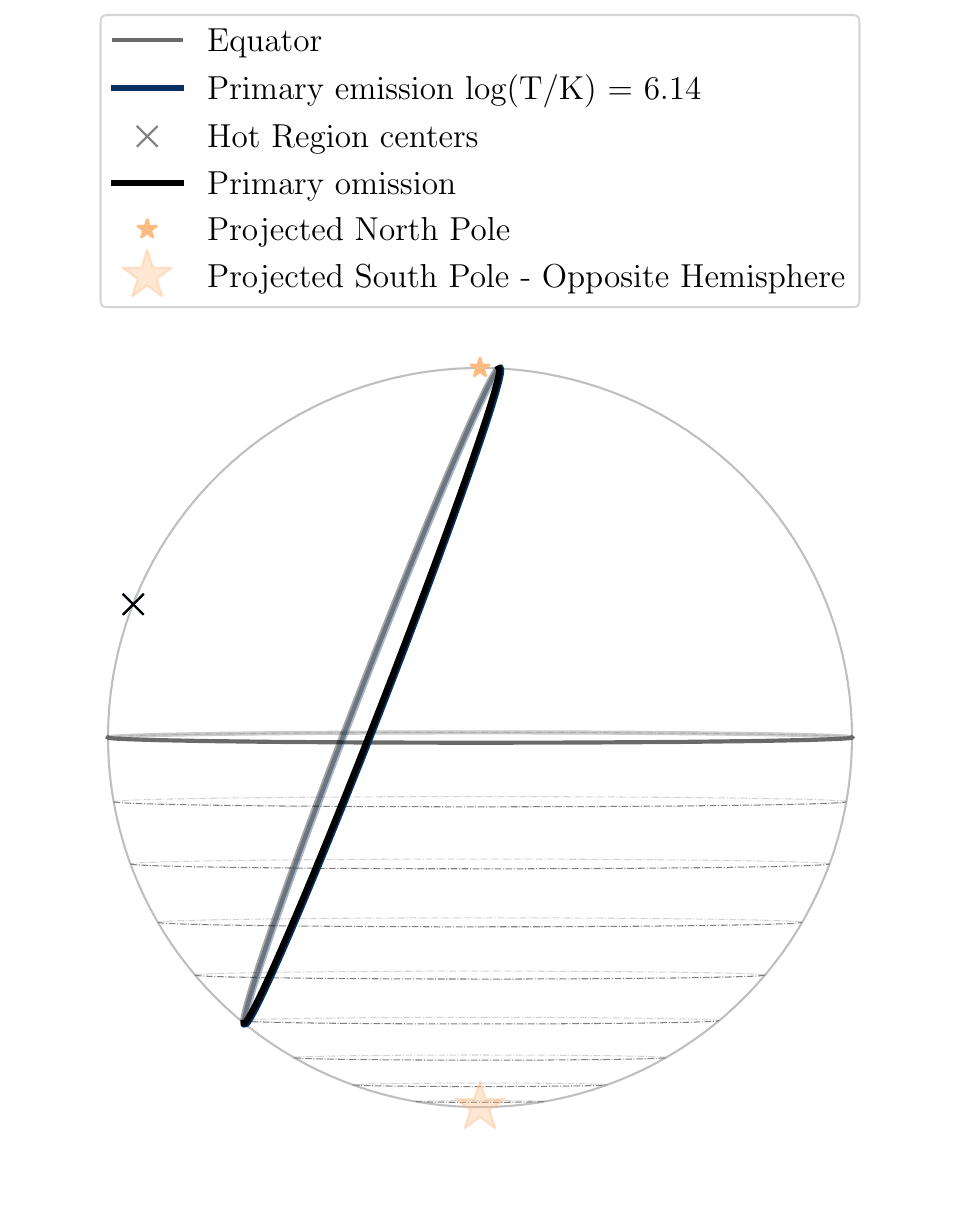}
    \caption{The \CST maximum likelihood geometry as seen from Earth, without gravitational effects for clarity. The CST forms a ring going from pole to pole, whose width is thinner than the lines on this plot. We deem this non-headline model unrealistic, although it underlines the need for better physics-informed priors on the hot spot geometry.}
    \label{fig:CST}
\end{figure}

While processing the \STCST run, we discovered that the posteriors were mostly constraining the CST hot spot, while the ST spot was at low temperature, making its contribution to the pulse small. As a result we decided to test a \CST only geometry, which is equivalent to having a single ring emitting at the surface of the MSP. This resulted in an overall thin and large ring, going from pole to pole. The maximum likelihood geometry (Figure \ref{fig:CST}) has a ring width of $\sim 0.3\degree$, while the ring width posterior has a median at $\sim 0.7\degree$ and an upper 90th percentile of $\sim 2\degree$. This simpler geometric model provided a similar improvement of  evidence as \STCDT over our favored \STU geometry (see Table \ref{tab:modelComparison}). This geometry produces the peak of the pulse when the full ring is visible, and makes the unpulsed component when only part of the ring is visible.

However, the literature does not provide much support for such a thin ring going from pole to pole, with the absence of associated hot spots at the polar caps \citep{Gralla+16,Gralla+17,Lockhart+19,Huang26}. This geometry being favored occurs most likely as a consequence of the low count statistics of this source and the single broad peak in the pulse profile, which allows oversimplified models, such as CST only, to fit the data. Although the fit is worse (maximum log-likelihood is $\sim 2.7$ lower than for \STU), the lower number of parameters of \CST (three less than \STU) favors it when using the evidence for model comparison (log-evidence increases by $\sim 2.7$). Moreover, using directly the Bayesian evidence for model comparison assumes the models to be equally probable, otherwise the Bayesian evidences should be weighted by the prior belief. Although it is not possible to give quantitative values easily, the \STU model is preferred a priori given the current theoretical knowledge about hot spot formation \citep{Chen+20,Kalapotharakos+21,Petri22,Stefanou+23}.

Hence, this CST-only geometry does not provide mass-radius posteriors suitable for downstream EoS analysis. However, we deem this geometry still interesting to report, as it highlights the need for better physical constraints on the hot spots as input to our geometric priors. 

\subsection{Different atmosphere models}
\label{subsec:atmosphere_fit}

For our preferred \STU model, we tested two other NSX atmosphere models, different from the fully ionized hydrogen one that was used throughout all the other exploratory runs. These consist of a fully ionized helium model (He) and a partially ionized hydrogen model (Hp). All atmosphere models give comparable posteriors for all parameters. The posterior distributions of the geometric parameters are mostly unchanged, although the temperature of the primary spot changes by $\Delta \log (T_p \,\rm{[K]}) \simeq 0.07$. We note that the span of the effective surface gravity $\log_{10}(g \,[\rm{cm/s}^2])$ in the helium atmosphere table is smaller, and only goes up to 14.7, while both hydrogen atmosphere tables reach 15.0. This affects the inferred radius of the helium atmosphere run, whose prior-imposed lower limit is $\sim 500\,$m higher than for the hydrogen runs.\footnote{We note that this effect did not occur in the study of PSR~J2124$-$3358 from \citealt{Gonzalez-Caniulef+26} as the inferred compactness ($C\simeq0.23$) is much lower than for \jsf and these analyses were performed using common limits for the effective surface gravity for the hydrogen and helium atmosphere models.}

The respective maximum likelihoods and evidences of these runs are provided in Table~\ref{tab:modelComparison}. The slight increase in evidence for the helium atmosphere is in part due to the tighter effective surface gravity, restricting the prior and artificially increasing evidence. However, even with this small increase, all three evidences are comparable, meaning that none of these is favored by the data. Hence, we pursue our \STU headline run using the fully ionized hydrogen atmosphere model.

\subsection{Headline \STU}
\label{subsec:headline}

For the headline results, we chose to use the \STU model, which delivered similar, or better, evidence compared to more complex models (see Section~\ref{subsec:geometric_configs} for further explanations on this choice). As previously mentioned (Section \ref{subsec:sampling}), we increased the sampler settings to $4\times10^4$ LP and $0.03$ SE, while keeping the ET fixed (compared to $2\times10^4$ LP and $0.05$ SE for the exploratory runs). On the other hand, the X-PSI resolution settings were kept the same as for exploratory runs, based on a similar analysis as in the Appendix A of \cite{Mauviard+25}. Applying the same method here showed that higher reasonable X-PSI resolution settings did not provide any improvements in likelihood precision compared to the high resolution, remaining around $\Delta \ln L \sim0.1$ while increasing the computation time. This is due to the LR settings being high enough to provide an accurate description of ST hot spots, which are the simplest geometry we can have in X-PSI.

The headline run resulted in similar constraints as the exploratory \STU run (see Section \ref{subsec:ST-U}). The median radius decreased by $\sim40\,$m while the 68\% CI increased by $\sim60\,$m, mostly extending the high radius tail (see Figure~\ref{fig:MR_QUR}). We find again a geometry with one hot spot near the pole and one near the equator, with two degenerate modes. The case where only one ST is visible is still present (see Section \ref{subsec:degeneracies_T}), indicating that this part of the parameter space cannot be firmly excluded. However, the individual modes are now better split from this case than they were in the exploratory \STU run. This manifests itself with the absence of a secondary peak feature around $\phi_p \sim 0.75$ in the marginal primary phase distribution (see Appendix \ref{app:full_posterior}).

\begin{figure*}
    \centering
    \includegraphics[width=0.8\linewidth]{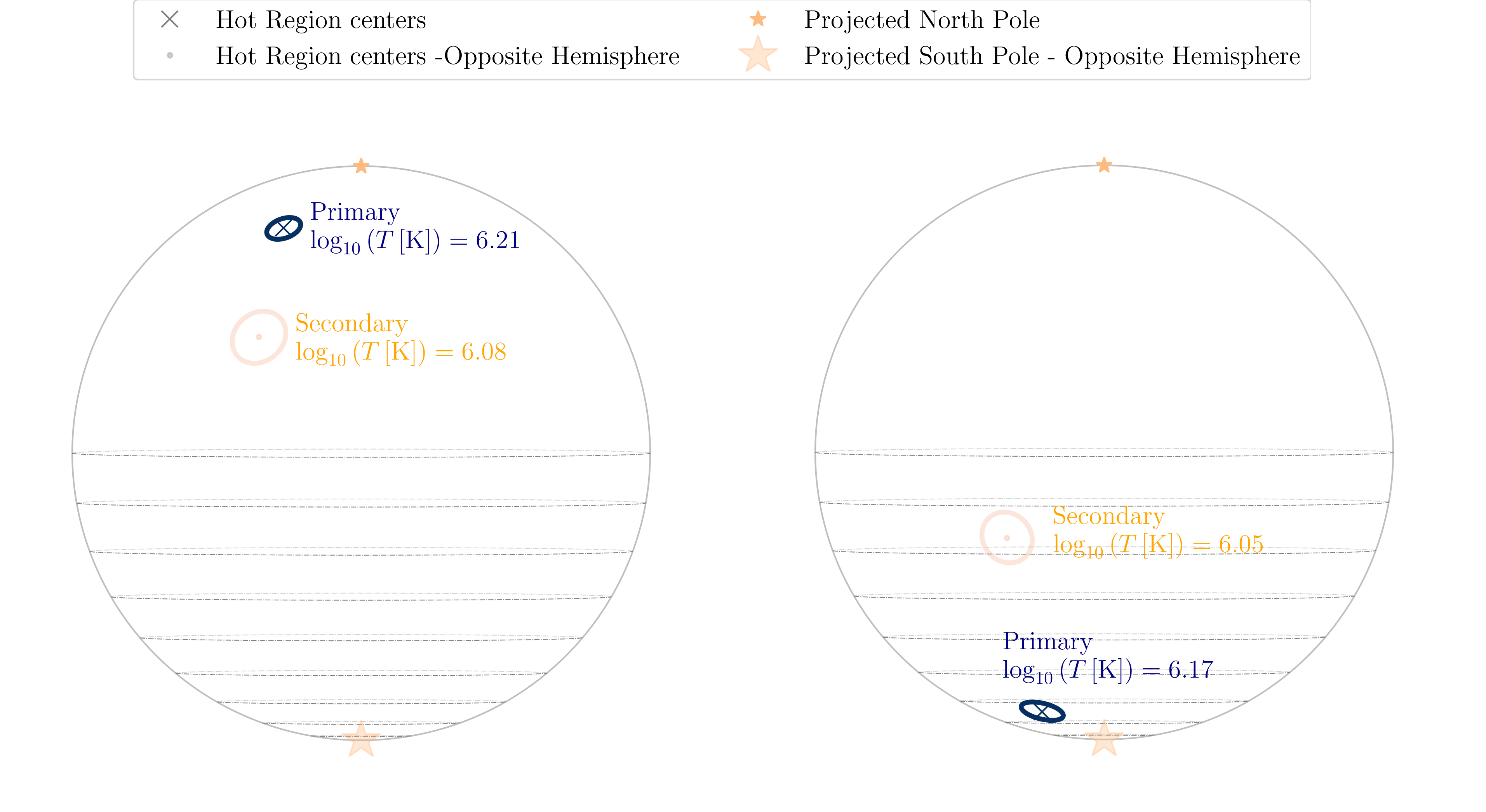}
    \caption{Representation of the best-fit geometries, as seen from Earth, for the two modes present in the headline \STU run. The pulsar is represented without gravitational or relativistic effects at phase zero of the data. The hotter spot is represented in blue, while the cooler one is in orange. These two modes are North/South mirror images due to degeneracy in the model. The secondary hot spots are located behind the star in both plots. Both geometries have similar likelihood. These representations do not encompass the full posterior distributions of the geometric parameters.}
    \label{fig:BestFitGeometries}
\end{figure*}

For visualization purposes, the maximum likelihood geometries of each mode are shown in Figure \ref{fig:BestFitGeometries}, and are degenerate versions of each other. These are maximum likelihood representations of the geometry, and the colatitude distribution of the equatorial hot spot is centered on the equator and spans a large range of values ($\theta_s = 1.57 ^{+0.52}_{-0.53}\,$rad). The colatitude of the secondary spot is actually weakly anti-correlated with the colatitude of the primary, making the configuration where each are in opposite hemispheres slightly favored. The size of the secondary hot spot $\zeta_s=0.28^{+0.28}_{-0.13}\,$rad is also largely unconstrained. Hence, one must not consider only these best fit geometries but rather the whole distribution to infer properties of the magnetosphere.

\begin{figure*}[t]
    \centering
    \includegraphics[width=\linewidth]{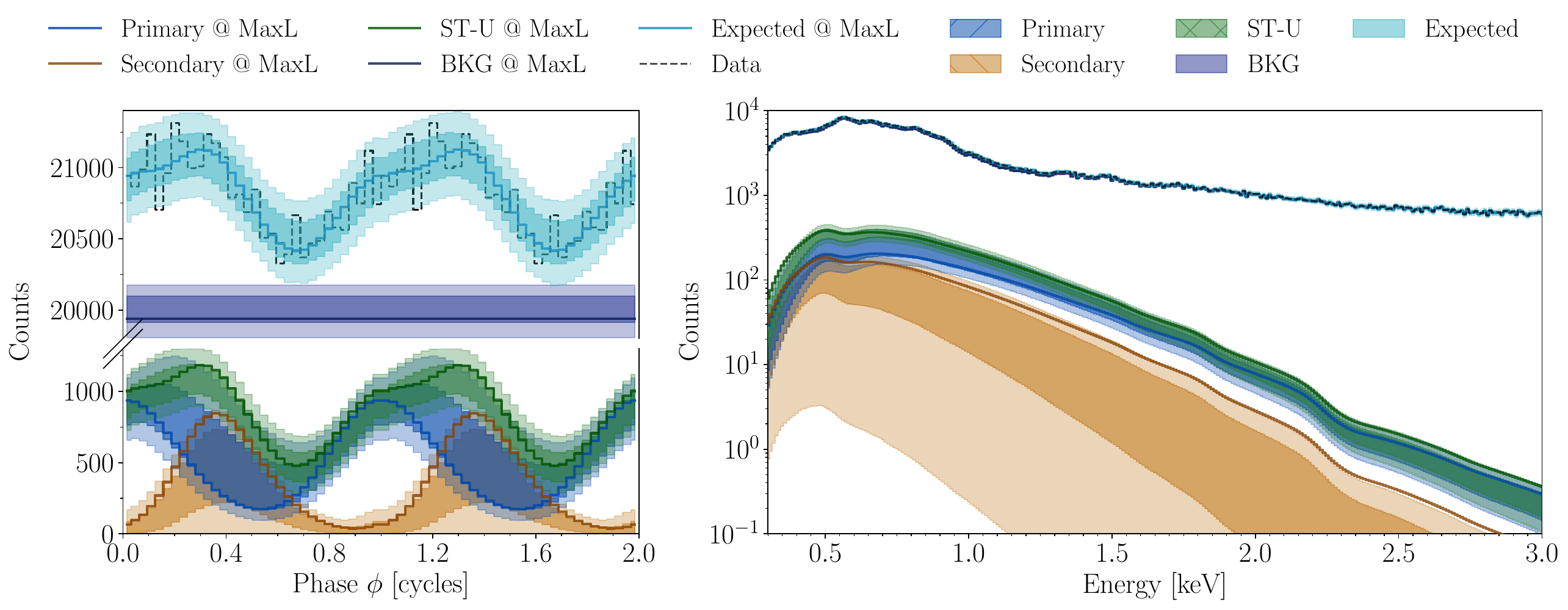}
    \caption{Posterior predictive distribution and maximum likelihood of the NICER observables for the headline \STU geometry: bolometric pulse on the left panel and phase-averaged spectrum on the right panel. The posterior predictive distribution represents the posterior probability distribution of the observables. The contributions of the individual model components are shown in blue for the primary ST (polar hot spot), orange for the secondary ST (equatorial hot spot) and purple for the inferred background (BKG). The combined model components are in green for ST-U, and in light blue for the total (expected, i.e., ST-U+BKG). The shaded regions represent the 68\% and 95\% CI of the individual model components (Primary, Secondary, ST-U and BKG) before the Poisson realization, since these components are not directly observable. The total expected model prediction, including the Poisson realization, is shown for comparison with the observed counts. The solid lines represent the maximum likelihood model and the black dashed line the data. The left panel has a discontinuous y-axis and the horizontal line at $\sim$20,000 counts represent the phase-independent maximum likelihood background. On the right panel, the background, expected distribution, and data all overlap.}
    \label{fig:posterior_pred}
\end{figure*}

The posterior predictive plots, which represents the posterior probability distribution of the observables, of both the NICER pulse profile and spectrum are shown in Figure \ref{fig:posterior_pred}. There, we can see that the NICER data consists mostly of background (95\%), which is indirectly constrained by the XMM and Chandra spectra of the pulsar. Moreover, we can see that the contribution of the secondary spot to the pulse is consistent with zero at $2\sigma$, which is associated with the case where only one of the hot spots is responsible for the pulse, the second being barely detectable (see discussion in Section \ref{subsec:degeneracies_T}).

This analysis yielded a posterior equatorial radius of \Req~$=10.06^{+1.25}_{-0.86}\,$km, which has the second widest 68\% CI (after \STUEL), overlapping with all the 68\% CI of the other models that were tested (see Section \ref{subsubsec:more_complex_models}). Moreover, the inferred neutral hydrogen column density \nh~$=(17\pm4)\times 10^{20} \,\rm{cm}^{-2}$ is correlated with the radius, and is consistent with previous X-ray spectroscopic analyses \citep{Pancrazi+12,Wolff+21} and neutral hydrogen sky maps \citep{hi4pi_2016}. The headline result also has a small asymmetry on the inferred mass posterior $M=1.937^{+0.012}_{-0.013}$. This is likely because at a given radius close to the compactness limit, we exclude more samples with slightly higher masses, producing the asymmetry.

\section{Discussion}
\label{sec:discussion}

\subsection{Low model dependency of the radius}
\label{subsec:low_model_dependency}
In our analysis, the source tends to have a compactness close to its upper limit of $C\lesssim0.33$, and hence a radius close to its lower limit ($R_{\rm eq} \gtrsim 8.5\,$km, depending on the mass, see Section \ref{subsec:prior}). This is true for all the tested models, regardless of the chosen geometry. Therefore, independently of the choice of geometric representation of the hot spot, \jsf has an equatorial radius centered on lower values ($\sim 9.5-10.5\,$km), with an upper 90th percentile of \Req$\lesssim12.08$\,km over all tested models (list in Table \ref{tab:modelComparison}), with the most extreme being the \STUEL geometry\footnote{CST is an exception, with a 90\% upper limit of 12.53 km, but as we have argued this model is unlikely to be physical in Section~\ref{subsubsec:CST}.}. Moreover, we also always tend to find one hot spot near the equator and one near the pole.

These can be understood from the source pulse and knowledge of the MSP inclination. On one hand, the source bolometric pulse has only one broad pulse, with a posterior predictive pulsed fraction of $\sim 58\pm7\%$ (see Figure \ref{fig:posterior_pred}), indicating the presence of a significant unpulsed component from the hot spots themselves. The pulsed fraction only weakly depends on the chosen geometric model, and is essentially determined by the XMM and Chandra source spectra, which both indirectly constrain the NICER source flux (see Section \ref{subsec:background}). On the other hand, \jsf has a high inclination ($89.4\degree$), which indicates that only spots near the pole can be visible throughout one full rotation of the MSP, and be responsible for unpulsed emission.

To accommodate both of these things, there is always one spot close to the pole to produce the unpulsed component. Another spot near the equator is then responsible for the bulk of the pulsed emission. However, a single hot spot does not provide an accurate enough description of both the unpulsed emission and skewed and broad bolometric pulse. Hence, the spot near the pole must also produce some pulsation by moving off the pole. To maintain the large observed unpulsed component, the light bending, and therefore the compactness, must increase. As the mass is fixed by the radio prior, the radius must decrease.

Adding a warm surface with the Elsewhere component took some contribution off of the unpulsed emission from the spot at the pole, which moved the radius towards slightly higher values ($\sim 200\,$m), further strengthening this interpretation. However, such a component is not sufficient to explain the totality of the unpulsed component. This interpretation is also valid for \josft, which features both a similar pulse, pulsed fraction and high inclination \citep{Mauviard+25}, and also has a radius centered around $\sim10\,$km. Hence, it is expected to have consistent constraints for all geometric models, all pushing toward a configuration with low radius and a hot spot near the pole.

Based on the XMM timing data from \cite{Pancrazi+12} which resulted in a pulsed fraction of $\sim100\%$, a previous study expected that applying PPM to \jsf would provide a lower limit on its radius \citep{Miller16}. Here, our findings with NICER data are opposite: we find an upper limit on the radius. Unlike previously expected, the unpulsed component is actually quite significant (pulsed fraction $\sim60\%$). We did not use the XMM timing data that this previous study was based on, which was obtained from a much shallower exposure ($\sim 20\,$ks, see Section~\ref{sec:XMMdata}) than the NICER data ($\sim 950\,$ks). Old calibration and contamination of the background by a neighboring source might have lead the background to be overestimated, therefore overestimating the pulsed fraction.

\subsection{Geometry}

Across all models tested, only a few general hot spot configurations arise. In particular, all the geometries with two hot spots had modes with one hot spot near the pole, and one near the equator. Such a configuration is expected given the data and viewing geometry (see previous section). Using double temperature hot spots (\STCDT and \STPDT) resulted in large (few tens of degrees) and warm ($\sim 10^{5.6}$\,K) supplementary components around \STU-like hot spots. Moreover, the inferred warm surface component in \STUEL had a temperature of $\log_{10}(T_{\rm EW}\,\rm{[K]})=5.56^{+0.06}_{-0.12}$, which is lower than the temperature of the supplementary components from \STCDT and \STPDT, but higher than what is found using ultraviolet data for \jofts \citep{Kargaltsev+04,Durant+12,Gonzalez-Caniulef+19,Qi+26}. This hints at the existence of a temperature gradient around the spots that would explain both the unpulsed emission and inferred temperature of the supplementary components.

The posterior assigns a 95\% probability to the primary spot (near the pole) having a smaller angular radius than the secondary spot (near the equator). The hot spot boundary matches the footprint of the last open filed line on the surface of the MSP. For an off-centered dipole with a similar surface magnetic field strength for both spots, the angle between the magnetic moment and the last open field line is larger at the equator than at the pole. Hence, it is expected for the hot spot near the equator to be larger than the one near the pole. Assuming a large scale dipolar field and the magnetic obliquity to be equal to hot spot colatitude, we can compute the maximum expected angular radius of the polar caps to be $\sim 0.5-0.6\,$ rad \citep{Gralla+17}. From magnetic flux conservation, it is expected that any additional multi-polar structure in the magnetosphere will reduce the area of the polar cap. Subsequently, the size of the polar caps in the dipolar case can be considered as an upper limit on the hot spot sizes \citep{Gralla+16,Gralla+17}. Removing samples with a larger hot spot than this limit in our headline \STU posterior, we find that the radius median slightly increases by $\sim 50\,$m. This not surprising as we mostly reject samples where only one hot spot is detectable, and this region of the parameter space tends to have a lower radius. Finally, we conclude that applying such cut on the spot size during inference could be interesting to better constrain the geometry, but would not change our results here.

The \CST-like geometry also arises for the \STCST and \STCDT models. These configurations only require a thin ring (few degrees width at most) going from pole to pole to reproduce the data, which has never been observed in previous analyses. Given the low $\sim 8.5\sigma$ significance of the pulse from \jsf, it is not surprising that this simple geometry could fit the data, although worse than the \STU geometry. This highlights the need for physical priors on the hot spot parameters, especially for sources with low count statistics like \jsf.

\subsection{Comparison to other MSPs}
\label{subsec:comparison}

\begin{figure}[t]
    \centering
    \includegraphics[width=\linewidth]{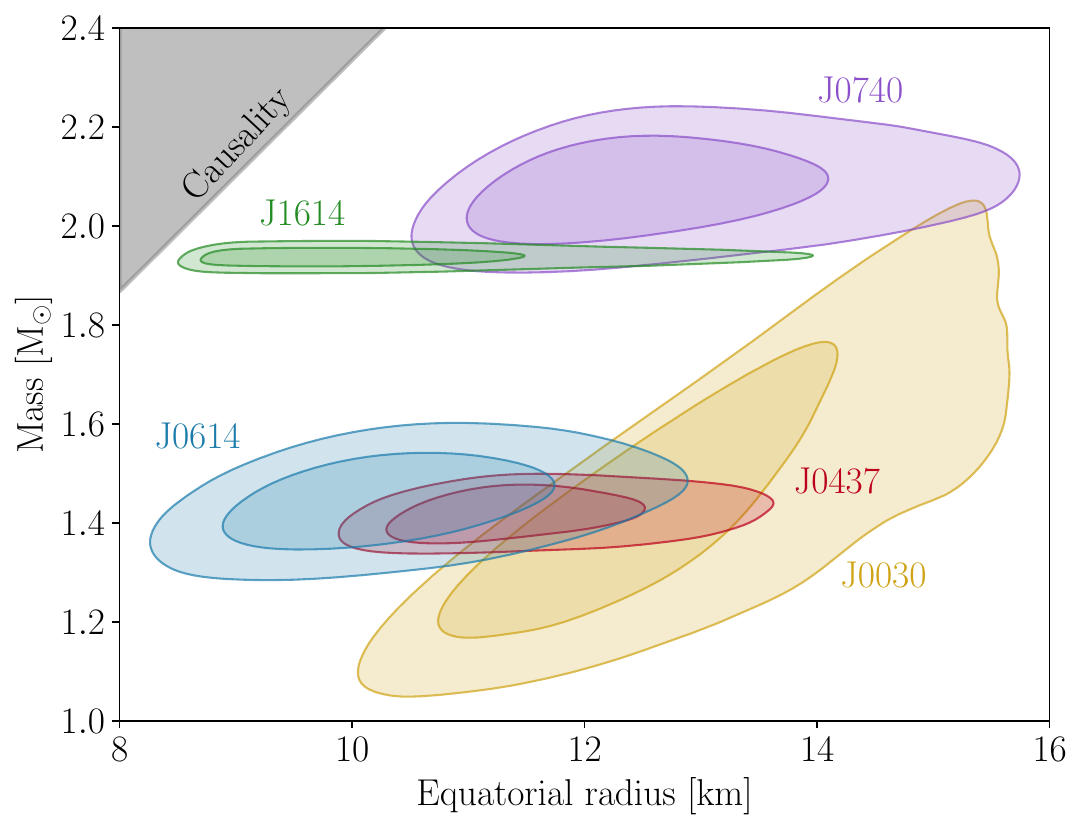}
    \caption{Mass-radius constraints obtained from MSPs with X-PSI. The \jsf posterior is the headline \STU model from this work, while other posteriors were extracted from \cite{Salmi+24_J0740} for \josfy, \cite{Choudhury+24_J0437} for \jofts, \cite{Mauviard+25} for \josft, and \cite{Kini+26} for \jdoty. The contours show the 68.3\% and 95.4\% credible regions for each source. The causality limit is plotted here for visualization purposes only, as the limit used during inference is slightly different and is applied to the polar radius (the reported values are from the equatorial radius).}
    \label{fig:MR_allSources}
\end{figure}

This PPM analysis of joint NICER, XMM and Chandra data from \jsf provides a second measured radius for a high mass MSP. The inferred distribution from this work sits at lower radius than that of the similar mass pulsar \josfy \citep{Salmi+24_J0740}, although they are still consistent (see Figure \ref{fig:MR_allSources}). Moreover, the lowest part of the inferred radius distribution is close to the causality limit \citep{Lattimer+07}, which is included in X-PSI as a upper compactness limit. 

This source shares important traits in common with \josft \citep{Mauviard+25}. Their broad and single peaked bolometric pulse profile and pulsed fraction are somewhat similar, although \jsf has lower flux, and they both have the same edge-on viewing geometry. Due to their similarity, the argument developed in Section \ref{subsec:low_model_dependency} also applies to \josft. Hence, it is expected that they both feature a rather consistent inferred radius across geometric models, leaning toward the lower edge of the prior range. This does not apply to \josfy, which is also seen edge on but has clear double peaked bolometric pulse profile.

Compared to other MSPs, \jsf has a better mass constraint, essentially due to its tight prior. Moreover, in this case the $M$--$R$ posteriors are close to the compactness limit, excluding some part of the $M$--$R$ space, and resulting in a prior-imposed lower limit on the radius and a final inferred mass that is slightly improved compared to the radio timing prior. However, the inferred compactness $C=0.28\pm0.03$ of this source has larger uncertainties than all the other MSPs. This is caused by the low source signal and total NICER exposure of this source, and results in larger radius uncertainties than what could have been achieved with a longer exposure. If the mass prior were to shift by $\pm 2\sigma$, we estimate that this would change the radius lower limit by $\pm 100\,$m, while keeping the 90\% upper limit unchanged. The effects of this new mass-radius posterior on EoS knowledge is presented in the companion article \citep{Mendes+26}.

Finally, although \jsf has a high and well contained mass, as well as its well-defined viewing angle, it is lacking in terms of count statistics compared to the other MSPs studied with NICER. Even with limited NICER exposure, we were able to obtain satisfactory radius posterior distributions, although the uncertainties can further decrease with deeper exposures of this source. It is hence a prime target for more X-ray observations, with either current or future observatories such as NewAthena \citep{Cruise+24} or eXTP \citep{Li+25_eXTP}. These will provide timing capabilities, alongside larger effective areas than NICER, which would be valuable to improve the radius constraint of \jsf. Moreover, their lower and constrained background will be useful to remove cross-calibration uncertainties from our analyses. 

\subsection{Caveats}

\subsubsection{QUR}

The updated fits to the QUR used only a limited number of simulated points at compactness comparable to the ones inferred with X-PSI ($C\sim0.3$, Figure~\ref{fig:QUR}). This is due to the masses used to compute these points, as they have been chosen to be within $\pm5\sigma$ of the prior mass. We use EoS with a semi-agnostic construction that rarely reaches such low radii at $\sim 2\,$\Msun, hence the high compactness part of the parameter space is limited in number of points. Even with these scarce points it appears that the trend can be extrapolated. Moreover, if we were not to use these updated QUR fits, we would have used the relations from \cite{AGM14}, which are only a few percent different in this high compactness range.

Our updated QUR fits only holds for \jsf, as it is tailored to its mass and spin frequency, while the version from \cite{AGM14} is more general. For \jsf, we found that using the updated results in a decrease of the inferred compactness (see Figure~\ref{fig:QUR}). This is caused by the slightly enhanced oblateness for a given compactness, making the required light bending effects less important. If the trend observed in Figure~\ref{fig:QUR} extends to other MSPs, we expect their inferred radius to increase such as it was the case for \jsf. 

\subsubsection{Convergence}
Our exploratory runs used MultiNest settings of $2\times10^4$ LP, $0.05$ SE and $0.1$ ET, while the headline \STU run used $4\times10^4$ LP, $0.03$ SE and $0.1$ ET as well. The headline run showed only minor changes from the exploratory run, but we still observed a small increase in radius 68\% CI by $\sim 60\,$m (3\% increase) compared to the exploratory \STU run, especially at high values. This hints that the same slow upper radius convergence as observed for \josfy may be present \citep{Salmi+24_J0740}. However, the temperature range was decreased and the temperature exchange degeneracy for \STU has been reduced (see Section~\ref{subsec:degeneracies_T}), providing better convergence for the same MultiNest parameters. Given the comparable data and MultiNest parameters as what was used for \josfy, as well as the marginal 3\% increase of the radius 68\% CI for the headline run, we do not expect substantial further broadening with enhanced sampling parameters.

\subsubsection{Inclination}

When using the radio timing estimate of orbital inclination for the spin inclination of the MSP, we assume that both are aligned. This is expected given the recycling process that has accelerated the NS spin up to millisecond period \citep{Guillemot+14,Lorange+26}. Moreover, previous radio and $\gamma-$ray light curve fitting of \jsf resulted in a spin inclination close to the orbital inclination ($i=78^{+12}_{-8}\,\degree$, \citealt{Johnson+14} and $i=89.1^{+0.5}_{-1.1}\,\degree$, \citealt{Lorange+26}). Nevertheless, we tried one \STU inference with a large uniform spin inclination prior, which resulted in one configuration consistent with the radio timing value of $\sim 89 \degree$, and another with $\sim 30\degree$ inclination. This new configuration is not consistent with the radio and $\gamma-$ray light curves, but does yield consistent radius with the preferred \STU model using the radio timing prior. Hence, using the orbital inclination from radio timing as a proxy for the spin inclination is likely justified for \jsf given the radio and $\gamma-$rays light curves, and does not introduce large uncertainties in our case.

\subsubsection{Spot Geometries}
For our favored \STU model, as well as for other more complex models, we have a possible configuration with only one visible hot spot near the pole fitting the data. This does not necessarily imply that there is only one hot spot on the surface of the MSP, and it can also be explained by a large interstellar absorption (\nh~$\sim10^{21}\,\rm{cm}^{-2}$) and cold hot spot ($\log_{10}(T_p \text{[K]})\lesssim5.6$). The low count statistics from \jsf makes many hot spot geometries possible. This results in degeneracies, and this single hot spot configuration favoring lower radii cannot be completely excluded. More exposure of \jsf will be valuable to better constrain the likely existence of the second hot spot. If the single visible hot spot configuration were to be firmly excluded with more data, or with physical hot spot prior, we can expect an increase in the inferred radius.

As was the case for other MSPs analyzed with PPM, we do not retrieve an antipodal hot spot configuration. Instead, we find a hot spot close to the pole and another close to the equator, much like for \josft \citep{Mauviard+25}. This likely comes from the similarity of \jsf to \josft in viewing angle geometry and pulse profile. \josfy is also seen edge-on but features two separate peaks, making the presence of a hot spot near the pole less likely \citep{Riley+21,Miller+21,Salmi+24_J0740,Dittmann+24}. In our case, we find a large deviation from antipodal spots, with an inferred deviation from antipodality of $77\pm28\degree$.

Our analyses are always somewhat dependent on the choice of the hot spot complexity used for reproducing the data, resulting in systematic uncertainties that are hard to quantify. In particular here, we find an interesting \CST geometry (see Section \ref{subsubsec:CST}) that is largely different from the polar and equatorial hot spot found across other analyses. This highlights the need for better geometry prior, as all our analyses so far have been completely agnostic to hot spot physics determining the shape and temperature distribution of the spots. Having informed geometric priors will prove to be of great help to get better mass-radius constraints for many sources, such as for \jsf or PSR~J1231$-$1411 \citep{Salmi+24_J1231}. However, we find that the inferred geometry and radius have actually little dependence on the hot spot complexity (see Section \ref{subsec:low_model_dependency}). The upper 90th percentile of the radius being 12.08 km over physical models shows that, regardless of the chosen model, the data tend to favor low radii. 

\section{Conclusion}
\label{sec:conclusion}

In this work, we have analyzed the X-ray pulsed emission of the massive MSP \jsflong using PPM. This source has high and well-constrained mass and inclination priors from radio timing (respectively $M=1.937\pm0.014\,$\Msun and  $\sin(i) = 0.9999019 \pm 0.0000037$). By jointly fitting NICER, XMM and Chandra data, using nested sampling, we inferred a consistent geometry and radius across all tested hot spot complexities. Our favored \STU model has a final constraint on mass of $M=1.937^{+0.012}_{-0.013}\,$\Msun and equatorial radius of $R_{\rm eq}=10.06^{+1.25}_{-0.87}$\,km. These are compatible with the previous measurement of \josfy, a MSP of similar mass, although \jsf leans toward lower radius values.

For this endeavor, we developed and used updated fits to the QUR tailored to \jsf spin frequency and mass range. We also tested the effect of using jointly Chandra and XMM, which individually give slightly different results. For the reported headline results, we took into account the cross-calibration uncertainties between them and use both data sets. This headline considers a fully ionized hydrogen atmosphere model. Moreover, we investigated the possible use of two different atmospheric composition: one with partially ionized hydrogen, and another with fully ionized helium. Both do not provide any significant improvement from the fully ionized hydrogen model. We also tested relaxing the prior on the spin inclination, which yielded two configurations, one of which constrains it to be in a similar range as the observed orbital inclination that we usually use as a proxy. Although a power law component was previously reported \citep{Pancrazi+12}, we did not find evidences for its existence in both spectroscopic and PPM analyses. 

The inferred geometry is composed of two circular hot spots, one near the pole and one near the equator, deviating from the centered dipole paradigm. There is a degeneracy on the pole where the primary hot spot sits, because the source is seen edge-on. Using different geometries, we have found hints of possible temperature gradients and warm surface emission. Another configuration with only the hot spot near the pole being visible and responsible for most of the pulsation is also possible, although disfavored by log-likelihood (difference in log-likelihood of $\sim 10$).

We propose that the rather low inferred radius of this source, but also that of \josft, are a consequence of their edge-on viewing geometry associated with the $\lesssim60\%$ pulsed fraction, which is mostly model independent. Together, they provide a lower limit on the compactness, and hence an upper limit on the radius. Using two hot spots, a high compactness $\sim 0.3$ star is required to explain both the broad and skewed shape of the bolometric pulse, and the large unpulsed component. We also show that an additive warm surface component would not be sufficient on its own to explain the low pulsed fraction. From such considerations, both this source and \josft \citep{Mauviard+25} are expected to yield rather low radii, independently of the used model. We find such trend, with an upper 90th percentile radius of 12.08 km across all tested physical geometries and atmosphere models.

Finally, these results were obtained with low count statistics NICER data, and constraints on the radius were made possible thanks to the high and tight mass and inclination priors. Deeper observations of this source, with either current of future observatories, will be valuable to further constrain the radius, and subsequently, the EoS of cold dense matter. 

%% Please use the acknowledgment and contribution environments. This will 
%% be anonomyized when the "anonymous" style option is used. 
\begin{acknowledgments}
The authors kindly thank Alexander Philippov and Simon Durpourqué for their insightful help. This study has been partially supported through the grant EUR TESS N°ANR-18-EURE-0018 in the framework of the Programme des Investissements d'Avenir. This work was performed using HPC resources from CALMIP (Grant 2016-19056). We acknowledge NWO for providing access to Snellius, hosted by SURF through the Computing Time on National Computer Facilities call for proposals. Part of this work has been supported by the CEFIPRA grant IFC/F5904-B/2018, ANR-20-CE31-0010 (MORPHER) and ANR-25-CE31-7901-01 (DENSER). All researchers from IRAP acknowledge the support of the CNES. M.H., Y.K. and A.L.W. acknowledge support from the NWO grant ENW-XL OCENW.XL21.XL21.038 \textit{Probing the phase diagram of Quantum Chromodynamics} (PI: Watts). T.S. acknowledges support by the Research Council of Finland grant No. 368807 and the Centre of Excellence in Neutron-Star Physics (project 374063). Work at NRL is supported by NASA. This work has relied on NASA's Astrophysics Data System (ADS) bibliographic services and the ArXiv. 
\end{acknowledgments}

\begin{contribution}
LM was responsible for the technical work, writing and submitting the manuscript.
\end{contribution}

\facilities{NICER, XMM-Newton, Chandra X-ray Observatory}

\software{ArviZ~\citep{arviz}, 
         Cython~\citep{cython2011},
         fgivenx~\citep{fgivenx},
         GetDist~\citep[][\url{https://github.com/cmbant/getdist}]{Lewis19},
         GNU~Scientific~Library~\citep[GSL;][]{Gough:2009},
         HEASoft \citep{HEASoft},
         IPython~\citep{IPython2007},
         Jupyter~\citep{Kluyver:2016aa},
         Matplotlib~\citep{Hunter:2007,matplotlibv2},
         MPI~\citep{MPI},
         MPI for Python~\citep{mpi4py},
         \textsc{MultiNest}~\citep{MultiNest_2009},
         nestcheck~\citep{higson2018nestcheck,higson2018sampling, higson2019diagnostic},
         NumPy~\citep{Numpy2011},
         OpenMP~\citep{openmp},
         \textsc{PyMultiNest}~\citep{PyMultiNest},
         Python/C~language~\citep{python2007},
         SAS \citep{XMM_SAS}
         SciPy~\citep{Scipy},
         X-PSI~\texttt{v3.3.0} (\url{https://github.com/xpsi-group/xpsi}; \citealt{xpsi})}


\appendix

\section{On the presence of a spectral power law component}
\label{app:powerlaw}

Previous works on \jsf reported a nonthermal component that could be modeled with a power law \citep{Pancrazi+12,Wolff+21}. Interestingly, the neighbor source (see Section \ref{sec:data}) also exhibits a hard power law-like spectrum. While \cite{Pancrazi+12} did not exclude this source from the extraction region, \cite{Wolff+21} carefully selected the extraction region to avoid it. However, in the latter case the power law is essentially unconstrained and they enforced a fixed power law index value of $1.6$. Moreover, the data we extracted reach the background level at $\sim 2\,$keV (see Figure \ref{fig:EPIC_Chandra_data}), hence, the evidence for this power law components is weak. But including a power law component can affect the inferred radius \citep{Qi+26,Miller+26,Guver+26}. This is expected, since a power law model is purely phenomenological and does not depend on the mass and radius of the source such that a smaller part of the signal is attributed to the thermal emission of the hot spots, which informs on mass and radius. Hence, it is important to further investigate the presence of a powerlaw in \jsf.

To do so, we performed joint time-averaged analyses on the XMM and Chandra data presented in Section \ref{sec:data}, as well as with the extracted spectra used in \cite{Pancrazi+12}. Using \texttt{jaxspec} \citep{Jaxspec}, we jointly fitted all the data with channels corresponding to the 0.3$-$10.0 keV range. We tested the \texttt{Tbabs}*\texttt{blackbody} (\textsc{bb}), \texttt{Tbabs}*(\texttt{blackbody}+\texttt{powerlaw}) (\textsc{bb+pl}), \texttt{Tbabs}*\texttt{nsatmos} (\textsc{ns}), and \texttt{Tbabs}*(\texttt{nsatmos}+\texttt{powerlaw}) (\textsc{ns+pl}) models. We did not include models with two thermal components, as a possible second component is mostly unconstrained.

\begin{figure}[!ht]
    \centering
    \includegraphics[width=0.95\linewidth]{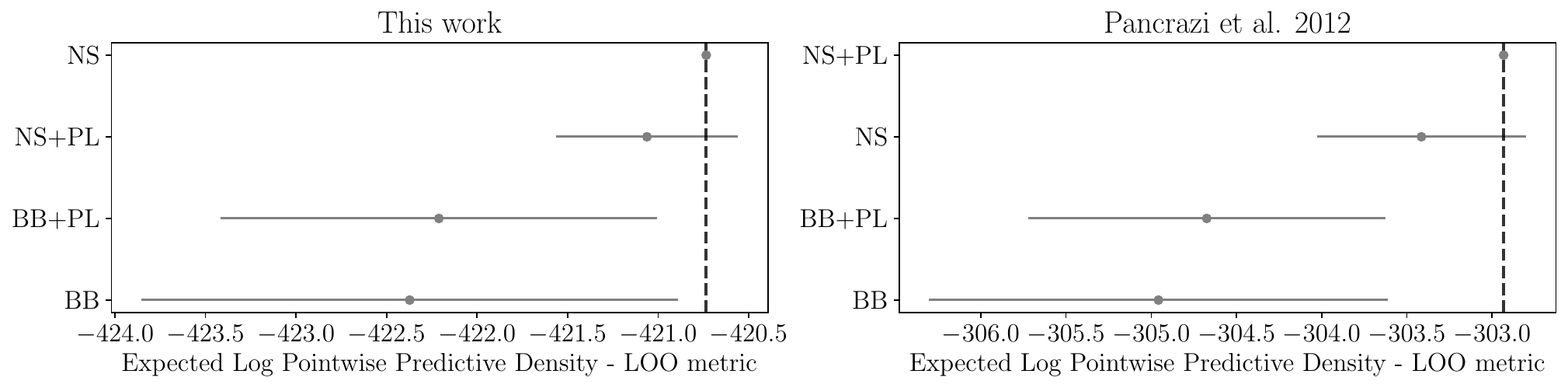}
    \caption{Model comparison plots using the Leave-One-Out (LOO) metric. Models are ordered by preference, with the better model on top. The gray horizontal lines provide the uncertainty of the difference of the LOO metric to the preferred topmost model. The dashed vertical line shows the LOO value for the preferred model. Left panel shows this comparison using the XMM and Chandra data curated in Section \ref{sec:data}, while the right panel shows the results using the data from \cite{Pancrazi+12}.}
    \label{fig:compare_PL}
\end{figure}

All the models managed to reproduce the data for both data sets. Using the arviz package \citep{arviz}, we performed Bayesian model comparison, whose results are provided in Figure~\ref{fig:compare_PL}. Overall, the models including a blackbody were disfavored compared to the ones with the atmosphere model. For our dataset, the \textsc{ns} model is slightly favored compared to \textsc{ns+pl}, while for the \cite{Pancrazi+12} data we find that the \textsc{ns+pl} model is slightly favored. This hints that the detected power law was likely not originating from the source, but rather due to contamination from the nearby source or the old calibration files.

Moreover, the normalization of the power law component in the NS+PL model with our dataset was on the lower edge of the prior, being consistent with zero ($\le 6 \times 10^{-7}$ photons/keV/cm$^2$/s at 90\% credibility). Using the XMM data, we find that fitting the neighbor source with an absorbed power law (\texttt{Tbabs}*\texttt{powerlaw}) results in a constrained norm of $(4.1\pm0.5) \times 10^{-6}$ photons/keV/cm$^2$/s. Considering that $\sim2\%$ of the flux from this source falls in the extracted region of \jsf (see Section \ref{sec:XMMdata}), this implies an estimated norm of $ \sim 10^{-7}$ photons/keV/cm$^2$/s, consistent with the contribution of the power law component in our spectral analysis of \jsf. All of the data, results and reproduction methods for these analysis is available in our Zenodo repository \citep{zenodo}.

We conclude that there is no significant power law component in our time-averaged data. However, such a component could still appear in the phase-resolved signal from NICER. Hence, we also added a power law component in our PPM analyses. This also resulted in non-detection of this component, as the models including a power law, pulsed or unpulsed, was disfavored by Bayesian evidence and the norm was inferred to have an upper limit of $\le 2.1 \times 10^{-6}$ photons/keV/cm$^2$/s (at 90\% credibility, see Section \ref{sec:diff_models_xpsi}).

\section{Automatic truncation of the Chandra response with specextract}
\label{app:Chandra}

The Chandra CIAO \texttt{specextract} task, used to extract the spectra, runs by default with the option \texttt{energy=0.30:11.0:0.01}. This indicates that the response of the instrument will be computed for incoming photons of energy spanning from 0.30 keV to 11.0 keV in steps of 0.01 keV.  In doing so, \texttt{specextract} truncates the ARF at the lower energy bound given by this option, i.e., 0.3~keV by default. It is possible to decrease this bound down to 0.2 keV by forcing the option to be \texttt{energy=0.20:11.0:0.01}. Here, we demonstrate that this option must absolutely be used to avoid truncating the response, which biases the results when fitting Chandra-ACIS data in an energy range starting at 0.3~keV.  

Let's consider that we fit a model to Chandra-ACIS data, including channels starting at 0.3 keV. Because of the spectral redistribution (encoded in the RMF), some photons with incident energies from the range 0.2--0.3~keV are expected to be detected by the camera in the instrument channels corresponding to energies 0.3--0.4~keV. The contribution of these photons is not negligible because of a transition edge at 0.27~keV in the ARF, making the effective area reach $\sim180\,\rm{cm}^2$ at that energy. However, this contribution is wrongfully ignored when using a response that has been truncated at 0.3~keV because of the default \texttt{specextract} option \texttt{energy=0.30:11.0:0.01}.  Because it effectively neglects the redistribution of photons with energy $<0.3$\,keV into channels corresponding to $>0.3$\,keV, this results in a lack of modeled counts at lower energies when fitting the data, and therefore in a bias on the inferred parameters.

\begin{figure}[!ht]
    \centering
    \includegraphics[width=0.49\linewidth]{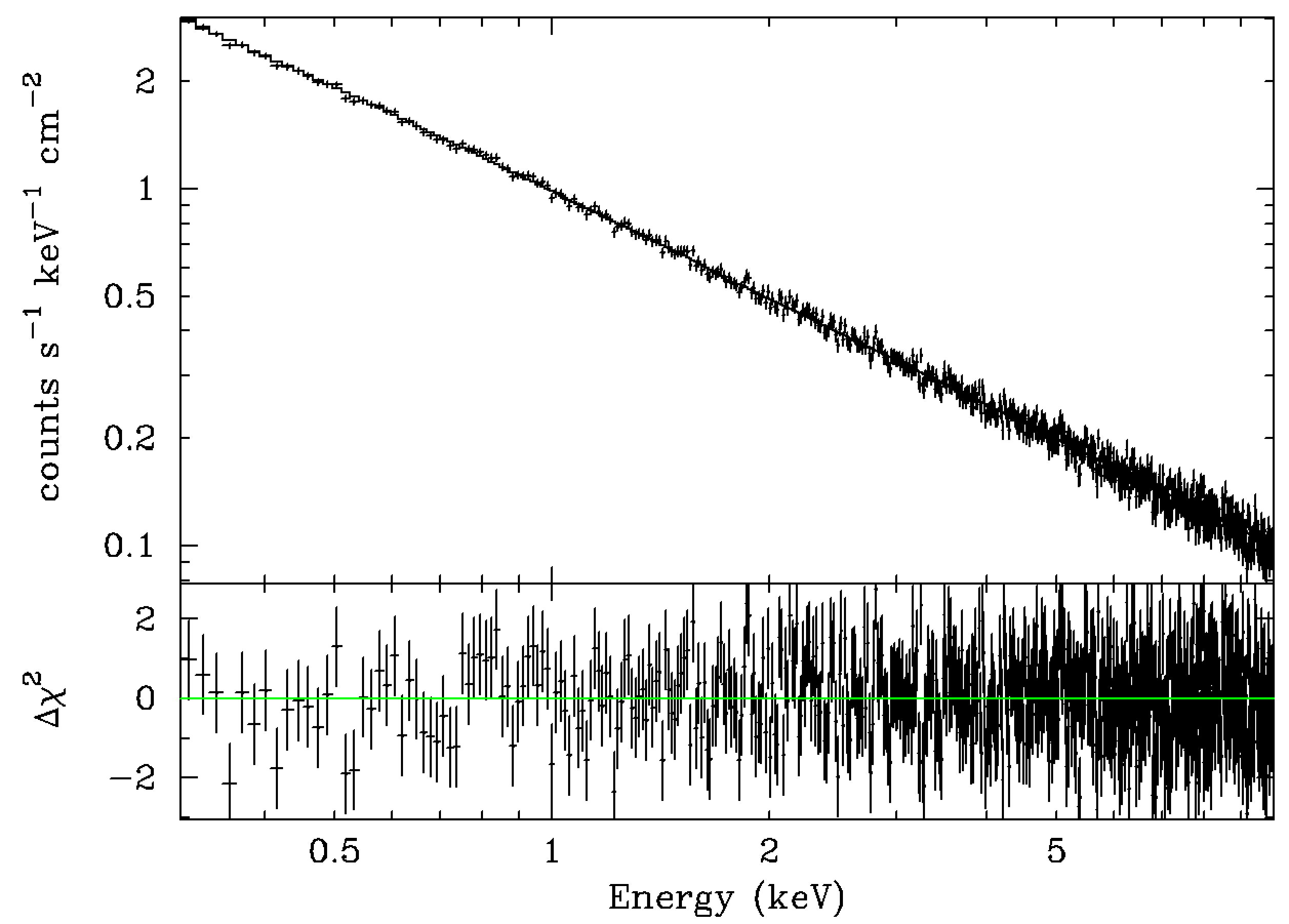}
    \includegraphics[width=0.49\linewidth]{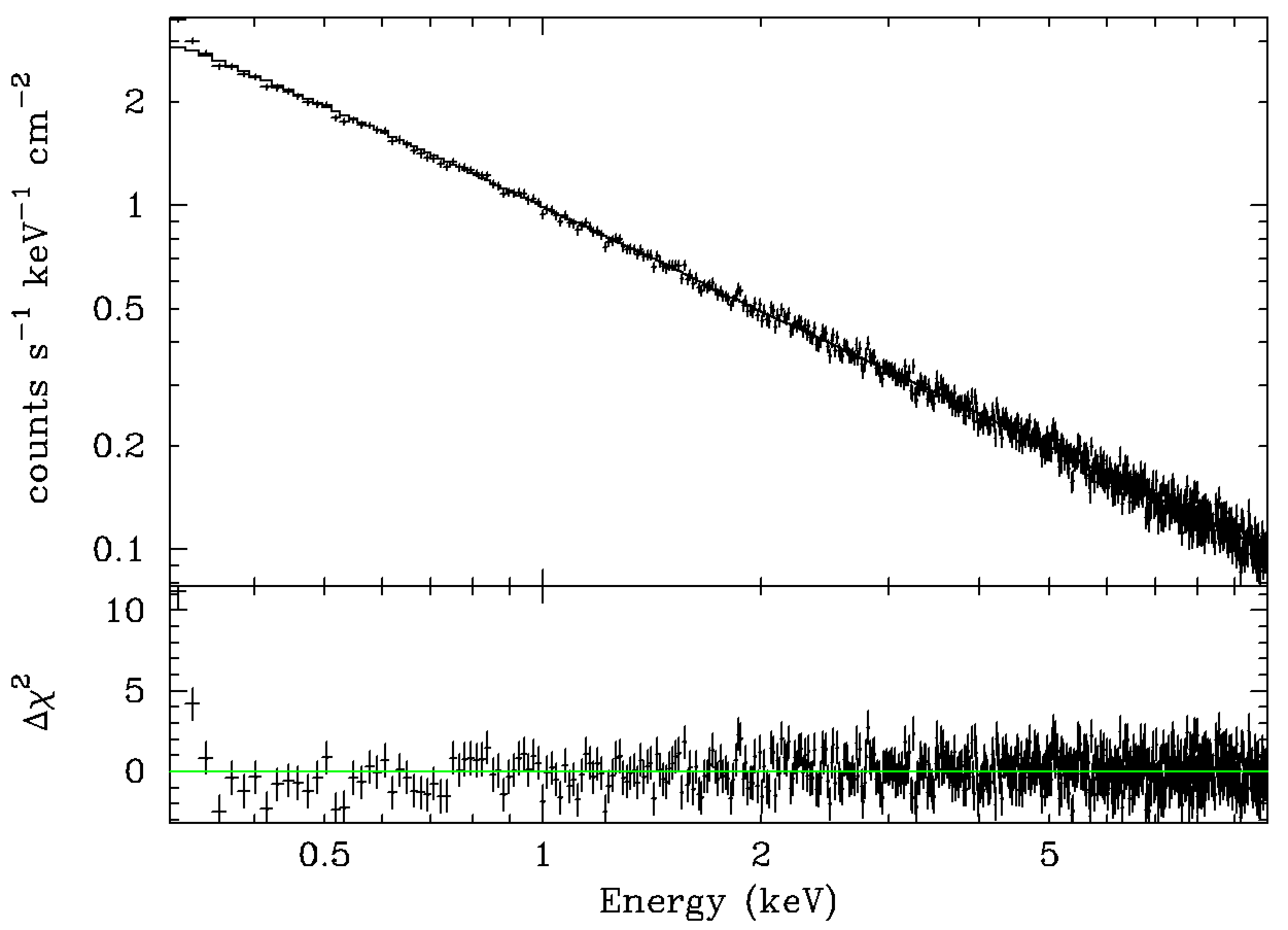}    
    \caption{Mock \textsc{pl} data alongside best fit model and residuals. Left panel: The spectrum was simulated and fitted with the non-truncated matrix. Right panel: The spectrum was simulated with the non-truncated matrix, but fitted with the truncated matrix with the default \texttt{specextract} option, showing a count deficit at low energies.}
    \label{fig:Chandra_response}
\end{figure}

To show the effect of using the truncated response with the default option to fit data down to 0.3~keV, we simulated a power law \textsc{pl} and an absorbed black body \textsc{bb} in XSPEC using a full (i.e., not truncated) response created with \texttt{energy=0.20:11.0:0.01}. We then fit these mock spectra in XSPEC with three responses created with the lower energy bound of 0.2~keV (the one used for the simulated spectrum), 0.25 or 0.3 keV. The best-fit parameters obtained when using the first two responses are consistent with the input values, while when using the response created with the 0.3~keV lower bound, the best-fit parameters are inconsistent with the input values (see Table~\ref{tab:Chandra_response}). When comparing the spectrum and model, we notice an drop in the model at lower energies when fitting with the 0.3~keV response (see Figure~\ref{fig:Chandra_response} for the PL). This count deficit is caused by the trimming of the response obtained using the default \texttt{specextract} task. It causes the model to be $\sim10\sigma$ away from the data points and the best fit PL index to be about $4\sigma$ different than 1.0.

\begin{table}[!ht]
    \caption{Recapitulative table of the fits proceeded using trimmed response. All errors are at the 90\% confidence level. $^\star$ The errors were estimated with the \texttt{steppar} command as $\chi^2_{\nu}>2$, indicating a bad fit.}
    \hskip-0.75cm \begin{tabular}{c|cc|ccc}
        \hline \hline 
        Model & Parameters & Input values & \multicolumn{3}{c}{Fitted values with response trimmed down to}  \\
         & & & 0.20 keV & 0.25 keV & 0.30 keV \\
         \hline 
         \textsc{pl} & norm [photon/keV/cm$^2$/s] & 1.0 & $1.003\pm0.002$ & $1.003\pm0.002$ & $1.008\pm0.002$ \\
         & Photon index & 1.0 & $1.003\pm0.002$ & $1.003\pm0.002$ & $1.007\pm0.002$\\
         \hline 
         & \nh [$10^{20}\,\rm{cm}^{-2}$] & 1.0 & $1.02 \pm 0.09$ & $1.02 \pm 0.09$ & $0.42\pm0.09$$^\star$\\
         \textsc{bb} & Black-body norm & 0.05 & $0.0500\pm0.0005$ & $0.0500\pm0.0005$ & $0.0476\pm0.0004$$^\star$\\
         & temperature $k_BT \, \rm{[eV]}$ & 150 & $150.0\pm0.5$ & $150.0\pm0.5$ & $151.7\pm0.5$$^\star$\\
         \hline \hline 
    \end{tabular}
    \label{tab:Chandra_response}
\end{table}

Comparatively, the results for the responses trimmed at 0.25 and 0.2~keV are virtually identical due to the negligible redistribution of photons with energies $<0.25$\,keV into channels with energies $>0.3$~keV. A similar situation is observed for the absorbed blackbody, where the deficit of counts in the first few bins is compensated by an artificially small \nh, inconsistent with the input value. Conversely, the temperature is overestimated. Overall, because of the redistribution in energy, it is crucial to extract the Chandra response in an energy range that extends beyond the energy interval that will be used for spectral analysis. For an analysis energy range starting at 0.3~keV, the default option of \texttt{specextract} is incorrect and needs to be manually adjusted. 

\section{Full posterior distribution of the headline \STU}
\begin{figure}[!ht]
    \centering
    \includegraphics[width=\linewidth]{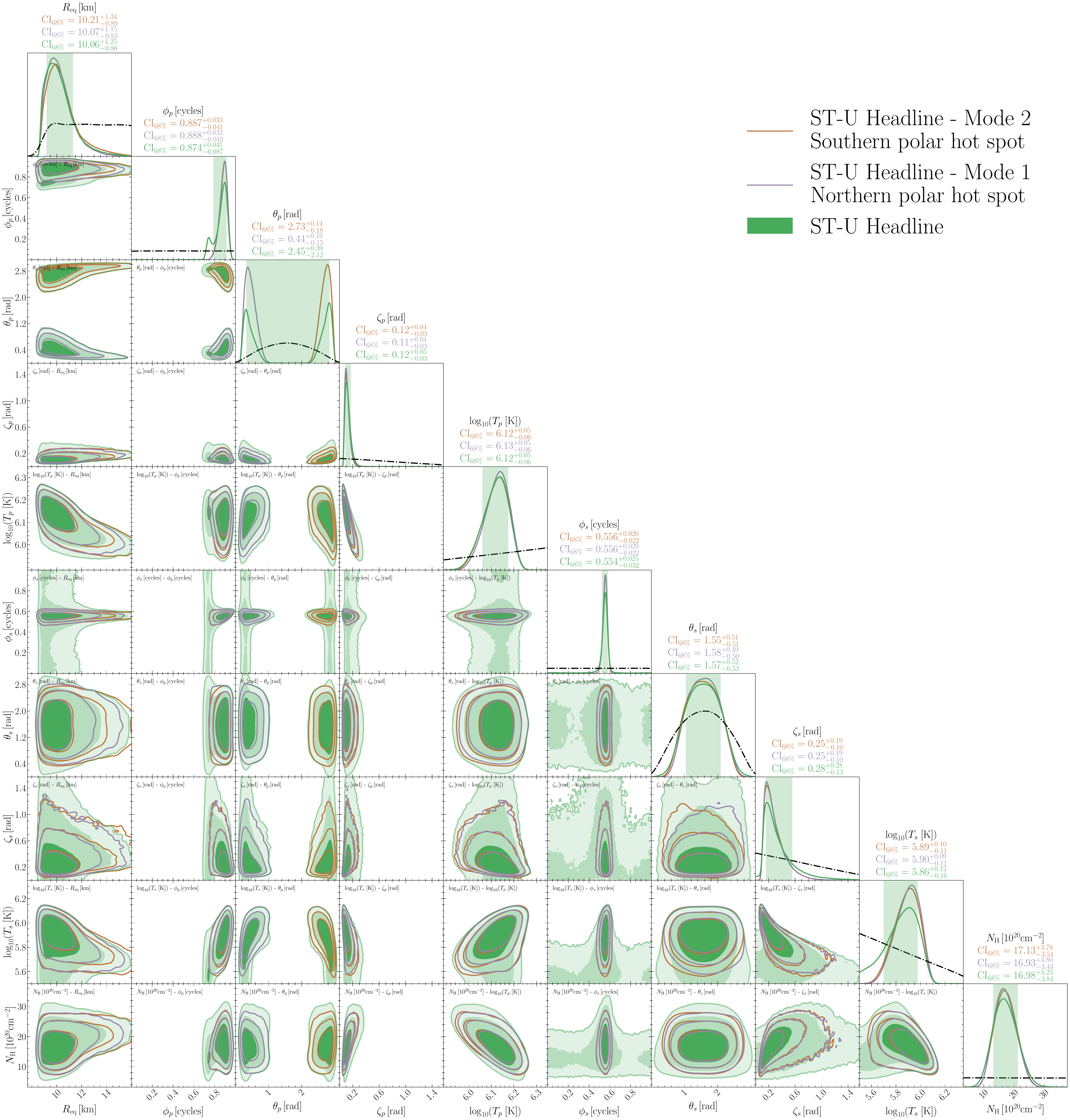}
    \caption{Posterior distribution of the radius, hot spot geometric parameters, temperatures, and hydrogen column density for the headline \STU run. Individual mode distributions (purple and orange) are plotted along with the full distribution (green). Mode 1 corresponds to the left plot from Figure~\ref{fig:BestFitGeometries}, and Mode 2 corresponds the right one. The dashed-dotted black lines on the diagonal plots show the marginalized prior distributions, shared across all runs. The shaded vertical bands show the $68.3\%$ CI. The contours in the 2D posteriors show the $68.3\%$, $95.4\%$, and $99.7\%$ credible regions, which are filled for the mode-averaged distribution only.}
    \label{fig:fullPoserior}
\end{figure}

The headline \STU run summary is given in Table~\ref{tab:headline_result}. The posterior distribution of a subset of parameters from the headline \STU run, with individual modes detailed, is given in Figure~\ref{fig:fullPoserior}. The figure of the full posterior distributions, including all 18 parameters, can be found in the Zenodo repository \citep{zenodo}. 

\label{app:full_posterior}  
\begin{table*}[!ht]
    \caption{Summary table for the headline \STU run.}
    \hskip-1.75cm \begin{tabular}{llllll}
    \hline \hline 
        Parameter & Description & Prior PDF & $\widehat{\rm CI}_{68\%}$ & $\widehat{\rm ML}$ & $\widehat{D}_{\rm KL}$ \\
        \hline
        $P$ [ms] & Coordinate spin period & fixed 1/317.38 ms & \\
        \hline
        $M$ [$M_\odot$] & Gravitational mass & $\mathcal{N}(1.937,0.014)$ & $1.937^{+0.012}_{-0.013}$ & 1.957 & 0.018 \\
       \Req [km] & Coordinate equatorial radius& $\mathcal{U}(6, 16)$ & $10.06^{+1.25}_{-0.87}$ & 9.83 & 0.944 \\
        $D$ [pc] & Distance to the source & $\mathcal{N}(678,42)$ & $689^{+34}_{-38}$ & 722 & 0.069 \\
        $\sin ( i )$& Sine of angle from line of sight to spin axis & $\mathcal{N}(0.999902,0.000004)$ & $0.999902^{+0.000003}_{-0.000003}$ & 0.999899 & 0.017 \\
        \hline
        & Compactness condition & $R_{\rm polar}/r_g(M) > 3$ & & &\\
        & Surface gravity condition & $\forall\theta,\ 13.7 \le \log_{10} g(\theta) \le 15.0$ & & &\\
        \hline
        $\phi_p$ [cycles]  & p region - initial phase & $\mathcal{U}(0,1)$ & $0.874^{+0.041}_{-0.087}$ & 0.927 & 2.218 \\
        $\theta_p$ [rad]  & p region - center colatitude & $\cos(\theta_p) \sim \mathcal{U}(-1,1)$ & $2.45^{+0.39}_{-2.12}$ & 0.66 & 2.208 \\
         $\zeta_p$ [rad]  & p region - angular radius & $\mathcal{U}(0,\pi / 2 )$ & $0.121^{+0.050}_{-0.033}$ & 0.06 & 2.640 \\
        $\log_{10}(T_p \text{[K]})$ & p region - effective temperature& $\mathcal{U}(5.5,6.5)$ & $6.123^{+0.051}_{-0.062}$ & 6.214 & 1.853 \\
        $\phi_s$ [cycles]  & s region - initial phase  $^{\rm a}$ & $\mathcal{U}(0,1)$ & $0.554^{+0.025}_{-0.032}$ & 0.563 & 2.534 \\
        $\theta_s$ [rad]  & s region - center colatitude& $\cos(\theta_s) \sim \mathcal{U}(-1,1)$ & $1.57^{+0.52}_{-0.53}$ & 1.17 & 0.184 \\
        $\zeta_s$ [rad]  & s region - angular radius& $\mathcal{U}(0,\pi / 2 )$ & $0.28^{+0.28}_{-0.13}$ & 0.10 & 0.636 \\
        $\log_{10}(T_s \text{[K]})$ & s region - effective temperature& $\mathcal{U}(5.5,6.5)$ & $5.86^{+0.12}_{-0.16}$ & 6.08 & 0.587 \\
        \hline
        & Non overlapping hot regions condition & & & &\\
        & Spot ordering based on temperature & $T_p \ge T_s$ & & &\\
        \hline
        $N_{\rm H} [10^{20} \text{cm}^{-2}]$ & Interstellar neutral hydrogen column density & $\mathcal{U}(0.001,100)$ & $16.98^{+4.23}_{-3.84}$ & 9.35 & 2.613 \\
        $\alpha_{\text{NICER}} $ & NICER effective area scaling factor & $\mathcal{N}(1.0,0.104)$ & $0.962^{+0.095}_{-0.085}$ & 1.033 & 0.105 \\
        $\alpha_{\text{EPN}} $ & PN effective area scaling factor & $\mathcal{N}(1.0,0.104)$ & $1.005^{+0.070}_{-0.069}$ & 1.024 & 0.180 \\
        $\alpha_{\text{EMOS1}} $ & MOS1 effective area scaling factor & $\mathcal{N}(1.0,0.104)$ & $1.018^{+0.071}_{-0.071}$ & 1.020 & 0.178 \\
        $\alpha_{\text{EMOS2}} $ & MOS2 effective area scaling factor & $\mathcal{N}(1.0,0.104)$ & $1.026^{+0.071}_{-0.071}$ & 1.075 & 0.195 \\
        $\alpha_{\text{ACIS}} $ & ACIS effective area scaling factor & $\mathcal{U}(0.5,1.5)^{\rm b}$ & $0.868^{+0.096}_{-0.089}$ & 0.877 & 1.398 \\
        \hline \hline
        & Sampling information & & & &\\
        \hline
        & Number of parameters: 18 & & & &\\
        & Number of live points: 40.000 & & & &\\
        & Sampling efficiency: 0.03 & & & &\\
        & Evidence tolerance: 0.1 & & & &\\
        & Multi-mode: On & & & &\\
        & Number of parameters for mode splitting: 12 & & & &\\
        & Likelihood evaluations: 34,187,066 & & & &\\
        & Estimated evidence: $\ln \mathcal{Z} = 54075.12 \pm 0.02$ & & & &\\
        & Computation time: 99,584 CPU hours & & & &\\
        & on AMD Rome 7H12 at 2.6 Ghz with 128 cores & & & &\\
        \hline \hline
    \end{tabular}
    \label{tab:headline_result}
        \begin{tablenotes}
        \item \textsc{Notes} - We report the prior PDFs, the median with the surrounding $68.3\%$ equally tailed CI $\widehat{\rm CI}_{68\%}$, the maximum likelihood \\ parameter vector $\widehat{\rm ML}$, and the Kullback–Leibler divergence $\widehat{D}_{\rm KL}$ in bits representing prior-to-posterior information gain.
        \item $^{\rm a}$ Unlike many previous analyses, the secondary spot phase is not shifted by $0.5$ cycles.
        \item $^{\rm b}$ This prior is different than the other cross-calibration factors due to a typo in all our run, and is more conservative. See Section \ref{subsec:prior}.
        \end{tablenotes}
\end{table*}

\clearpage

\bibliography{sample701}{}
\bibliographystyle{aasjournalv7}


\end{document}